\documentclass[prl, aps, 10pt, twocolumn, superscriptaddress, 
			showpacs, reprint, 
			]{revtex4-2}

\usepackage[T1]{fontenc}
\usepackage[utf8]{inputenc}
\usepackage[english]{babel}

\usepackage[top=2cm, bottom=2cm, 
			left=2cm, right=2cm,
			heightrounded]{geometry}

\usepackage{amsmath, amsthm, amssymb, mathrsfs, mathtools, dsfont}
\usepackage{physics, slashed, esint}
\usepackage[final]{microtype}
\usepackage{graphicx}
\usepackage{xcolor}
\usepackage{comment}
\usepackage{lipsum}

\usepackage[shortcuts]{extdash}

\makeatletter
\renewcommand*{\fnum@figure}{{\normalfont\bfseries \figurename~\thefigure}}
\renewcommand*{\@caption@fignum@sep}{\textbf{: }}
\makeatother

\usepackage{tikz}
\usepackage[compat=1.1.0]{tikz-feynman}

\usetikzlibrary{patterns, patterns.meta}
\usetikzlibrary{decorations.pathmorphing}
\usetikzlibrary{decorations.markings}
\usetikzlibrary{external}
\usetikzlibrary{positioning, arrows.meta}
\usetikzlibrary{calc, math}

\tikzfeynmanset{warn luatex=false}					
\pgfmathsetmacro\MathAxis{height("$\vcenter{}$")}		

\tikzset{with arrow/.style = {
   decoration={
     markings,
     mark=at position 0.5
          with {\arrow[xshift=0.8mm]{Stealth[width=1.1mm,length=1.6mm]}}
     },
   postaction=decorate}}

\tikzset{external photon/.style = {
	photon, 
	dash pattern = {on 0.5pt off 0.5pt}, 
	line width = 0.4pt}}

\tikzset{cross/.style = {
	path picture={ 
  		\draw[black]
  		(path picture bounding box.south east) -- (path picture bounding box.north west)
  		(path picture bounding box.south west) -- (path picture bounding box.north east);}}}
 
\tikzset{correlator background/.style = {
	pattern = {Dots[radius=0.15pt, angle=45, distance=0.7pt]}}}
	
\tikzset{external fermion/.style = {
	with arrow, 
	dash pattern = {on 0.5pt off 0.5pt}, 
	line width = 0.4pt}}

\newcommand{\dgauss}[2]{\mathbb{P}_{#1}(\dd{#2})}

\newcommand{\R}{\mathbb{R}}

\newcommand{\rff}{{\textsc{r}}}

\newcommand{\pfermi}{p^{\textup{\textsc{f}}}}
\newcommand{\loc}{\mathcal{L}}
\newcommand{\ren}{\mathcal{R}}
\newcommand{\dff}{\coloneq}

\newcommand{\absto}[2]{\abs*{#1}\vphantom{#1}^{\! #2}}
\renewcommand{\expval}[2]{\langle #2 \rangle}

\newtheoremstyle{thm}{}{}{}{}{\bfseries}{.}{.5em}{}

\theoremstyle{thm}

\newtheorem{theorem}{Theorem}

\newtheoremstyle{main}{}{}{}{}{\bfseries}{.}{.3em}{}

\usepackage{xurl}
\usepackage{hyperref}

\begin{document}

\title{Nonperturbative Chiral Anomaly Cancellation and Irrelevant Operators}

\author{Michele Bianchessi}
\affiliation{Università degli Studi Roma Tre, Mathematics and Physics Department}
\email{michele.bianchessi@uniroma3.it}

\author{Vieri Mastropietro}
\affiliation{Università degli Studi di Roma ``La Sapienza'', Physics Department}
\email{vieri.mastropietro@uniroma1.it}

\author{Marcello Porta}
\affiliation{SISSA, Mathematics Area}
\email{mporta@sissa.it}

\date{\today}

\begin{abstract} 
The Chiral anomaly cancellation is perturbatively well understood but
its nonperturbative validity in presence of a lattice has remained unproven.
We provide a rigorous nonperturbative proof of anomaly cancellation and its universality in a variation of the lattice Kogut-Susskind regularization
for the multiflavor Sommerfield model,
describing Dirac fermions interacting with a non-compact $\mathrm{U}(1)$ quantum vector field. We identify a novel and general
mechanism based on the convergence properties of the renormalized expansion and related to the one leading to universality in topological insulators. Irrelevant lattice operators, despite being suppressed along the RG flow, provide essential contributions to the anomaly cancellation. 
\end{abstract}

\maketitle

\paragraph{Introduction and main result.}

The 
chiral anomaly cancellation is a fundamental consistency requirement of the Standard Model.
Such a property is well understood at a perturbative level; yet, 
despite substantial progress, a nonperturbative proof is still lacking~\cite{Luscher_1999, Luscher_2000, Borrelli_1990, Kaplan_1992, Neuberger_2000, Suzuki_2000, Adams_2000, Golterman_2001, Kaplan_2024, Wang_Wen_2020, Golterman_Shamir_2024, Kikukawa_2019, Ginsprag_Wilson_1982, Grabowska_Dorota_Kaplan_2016, Grabowska_Dorota_Kaplan_2016_Regulator, Kadoh_Kikukawa_2008, Neuberger_1998}.
A finite lattice is the natural regularization for non-perturbative 
constructions, but only order by order results 
are known~\cite{Luscher_1999, Luscher_2000} while non perturbative ones are limited to non-renormalizable models~\cite{Mastropietro_Giuliani_Porta_2021, Mastropietro_2021_PRD_Electroweak_Anomaly}. 

In this letter we identify a new and general mechanism leading to the anomaly cancellation based on the lattice irrelevant terms in combination
with the convergence properties of the renormalized expansion.
It has similarities to the one leading to universality in presence of lattice and interactions of the edge conductance in topological insulators~\cite{Benfatto_Falco_Mastropietro_2010, Mastropietro_Porta_2022} and is technically implemented
by constructive Renormalization Group methods, see e.~g.~\cite{Douglas_2026, Kupiainen_2023} for reviews.
Rather than being mere corrections to scaling, lattice irrelevant operators provide essential contributions in recovering the anomaly-free continuum structure.
This is done in a two-dimensional renormalizable model, which 
provides a natural setting for investigating the anomaly cancellations. Previous analyses in two dimensions~\cite{Wang_Wen_2020, Georgi_Rawls_1971, Cheng_Seiberg_2023, Chatterjee_2025, Wang_Wen_2023, Berkowitz_2024, Kikukawa_2019_Decoupling, Thorngren_Preskill_Fidkowski_2026, Seifnashri_2026} were restricted to the non interacting case or were based on bosonization, whose extension to lattice regularizations is only approximate.

The model we consider is a variation of the lattice Kogut-Susskind regularization~\cite{Kogut_Susskind_1975, Susskind_1976} for the multiflavor Sommerfeld model~\cite{Georgi_Warner_2020}, describing
$n$ flavors of Dirac fermions in two dimensions. The fermions interact with a dynamical non-compact massive vector field, with coupling $e$, and a classical chiral gauge field, with couplings $e q_{c, \omega}$ ($c$ is a flavor label and $\omega = \pm$ denotes the chirality). 
The multiflavor Sommerfield model in the formal continuum limit is solvable by bosonization~\cite{Sommerfield_1964, Georgi_Warner_2020} and 
the solution is very sensitive to the choice of the regularization~\cite{Jackiw_1999, Hagen_1967}: on a lattice 
the model is not solvable.  
We therefore analyze the model using constructive Renormalization Group methods.
We prove that he anomaly vanishes under the condition $\sum_c q_{c,-}^2=\sum_c q_{c,+}^2$ up to subleading corrections. This is the same condition found in the continuum case (its $4d$ counterpart is $\sum_c q_{c,-}^3=\sum_c q_{c,+}^3$). 

\paragraph{The model.} 

We consider $n$ flavors of Dirac fermions described by the Grassmann variables $\psi^\pm_{c, x}$, where $c = 1, \dots, n$ and $x = (x_0, x_1)$. Here, $x_0$ is continuous and it belongs to $[-\beta/2, \beta/2]$, while $x_1$ is discrete and it belongs to a $1d$ lattice of step $a$ and length $L$. Antiperiodic boundary conditions are imposed on $\psi^\pm_{c, x}$. The expectation value of an observable
$\mathscr{O}(A,\psi)$ is
\begin{equation}
\label{eq:generating_functional}
\expval*{\mathscr{O}}= \frac{1}{\mathscr{Z}} \int \! \mathscr{D} \psi \mathscr{D} A \, e^{-S_0[A] - \sum_c S_{1, c}[A , \psi] + \int \! e 
\mathcal{J} \cdot Z} \mathscr{O}
\end{equation}
where $\mathscr{Z}$ is the normalization,
$S_0[A]$ is the free action of a massive vector boson $A_{\mu, x} \in \R$ whose propagator is $\hat{w}_{\mu \nu}(q) = \delta_{\mu \nu}(\absto{s(q)}{2} + M^2)^{-1}$ with $s(k) = (k_0, i(e^{-i a k_1} - 1)/a)$, and
\begin{multline}
\label{eq:lattice_action}
S_{1, c}[A, \psi] = \int_x \biggl[ \psi^+_{c, x} (\partial_0+ie A_{0, x} - t_c a^{-1}\cos a \pfermi_c) \psi^-_{c, x} \, + \\
- t_c \frac{\psi^+_{c, x} e^{-i a e A_{1, x}} \psi^-_{c, x + e_1}
+ \psi^+_{c, x + e_1} e^{ i a e A_{1, x}} \psi^-_{c, x}}{2a} \biggr]
\end{multline}
where $\int_x (\, \cdot \,) = a \int_{-\beta/2}^{\beta/2} \dd{x_0} \sum_{x_1} (\, \cdot \,)$ and $e_1= (0, a)$; $Z_\mu$ is a classical external field. The vacuum expectation value $\expval*{\mathscr{O}}$ is a series of connected correlations $\expval*{\mathscr{O}, \mathcal{J}, \dots}_0$ computed at $Z_\mu=0$.
The vector current $j_{\mu, c} = \partial S_{1, c}/\partial A_\mu$ carried by the flavor $c$ is exactly conserved, that is $s_\mu(p) \expval*{\hat{j}_{\mu, c, p}} = 0$; as a consequence, the longitudinal part of the gauge propagator $s_\mu(k) s_\nu(k)/(\absto{s(k)}{2} + M^2)$ does not contribute to $\expval*{\mathscr{O}}$. The chiral current is defined as $\mathcal{J}_\mu = j^\textsc{v}_\mu + j^\textsc{a}_\mu$, where $j^\textsc{v}_\mu = \sum_c Q^\textsc{v}_c j_{\mu, c}$ and $j^\textsc{a}_\mu = \sum_{c, \nu} Q^\textsc{a}_c Z^\textsc{a}_{\mu, c} (-i\epsilon_{\mu \nu}) j_{\nu, c}$. The parameters $Z^\textsc{a}_{\mu, c}$ are renormalization constants (see below)
and $Q^\textsc{v}_c= (q_{c,+}+q_{c,-})/2, Q^\textsc{a}_c= (q_{c,+}-q_{c,-})/2$. There is no lattice symmetry ensuring the conservation of the axial current.
Finally, note that the model is Hamiltonian, although it has been expressed in terms of a functional integral. 
At finite $L, \beta, a$, the functional integral is well-defined. Our aim is to remove the infrared cutoffs ($L, \beta \to +\infty$); the limit $a \to 0$ could be also taken~\cite{Fabbri_Renzi_Mastropietro_2026}, but here we are interested in keeping $a$ finite, as it is necessary in $4d$. For sake of simplicity, we shall take $a = M = 1$.

For every fixed $c$, the fermion propagator is equal to $\hat g_{c}(k) = (-ik_0 - t_c[\cos(ak_1) - \cos(a \pfermi_c)]/a)^{-1}$ and it is singular at $(0, \pm \pfermi_c)$. Close to these points, the dispersion relation is linear up to $\mathcal{O}(a \absto{k'}{2})$ corrections, where $k = (k'_0, \pm \pfermi_c + k'_1)$. The lattice model admits therefore an emergent description in terms of Weyl fermions $\psi^\pm_{c, \pm}$ with bare velocities $\abs*{\mathfrak{v}_c}\! = t_c \sin a \pfermi_c$ that can be reorganized into the Dirac fields $\psi_c = (\psi^-_{c, -}, \psi^-_{c, +}), \bar{\psi}_c = (\psi^+_{c, +}, \psi^+_{c, -})$. If we choose $\abs*{\mathfrak{v}_c}\! = 1$ for every $c$, the currents $j^\textsc{v}_\mu, j^\textsc{a}_\mu$ can be thought as the regularizations of the vector and the axial part of the emergent chiral current $\sum_{c, \omega} q_{c, \omega} \bar{\psi}_c \gamma^\mu[(1 - \omega \gamma^5)/2] \psi_c$, so~\eqref{eq:generating_functional} is the regularization of a theory of massless Dirac fermions with interaction $\int (A \cdot j  + Z \cdot \mathcal{J})$. The Kogut-Susskind formulation is recovered by letting $\pfermi_c = \pi/(2a)$ for every $c$.  

The anomaly is given by the axial-vector correlation function. It is therefore convenient to replace $e 
Z \cdot \mathcal{J}$ with a source term of the form $\int (-iJ^\textsc{a} \cdot j^\textsc{a} + \phi^+ \psi^- + \psi^+ \phi^-)$ and $A_\mu$ with $A_\mu+Q_c^\textsc{v} J^\textsc{v}_\mu/e$ in $S_{1, c}[A, \psi]$. The corresponding generating functional is called $W[J^\textsc{v}, J^\textsc{a},\phi]$; the axial-vector correlation function $\hat{\Pi}^{\mu \nu}(p) \equiv \expval*{\hat{j}^\textsc{a}_{\mu, p}, \hat{j}^\textsc{v}_{\nu, -p}}$, the vertex functions $\hat{\Gamma}^\sharp_{\mu, c}(k, p) \equiv \expval*{\hat{j}^\sharp_{\mu, p}, \hat{\psi}^-_{c, k}, \hat{\psi}^+_{c, k+p}}$ and the interacting two-point function $\hat{S}_c(k) \equiv \expval*{\hat{\psi}^-_{c, k}, \hat{\psi}^+_{c, k}}$ are then obtained by differentiating the generating functional with respect to $-iJ^\sharp, \phi^\pm$ (here, $\sharp = \textsc{a}, \textsc{v}$). 
The Ward Identities (WI)
\begin{align}
\label{eq:lattice_WI_current}
& is_\nu(p) \hat{\Pi}^{\mu \nu}(p) = 0 \\
\label{eq:lattice_WI_vertex}
& is^\mu(p) \hat{\Gamma}^\textsc{v}_{\mu, c}(k, p) = Q_c^\textsc{v} [\hat{S}_c(k+p) - \hat{S}_c(k)]
\end{align}
are valid and the normalization condition 
\begin{equation}
\label{eq:normalization_condition}
\hat{\Gamma}^\textsc{a}_{\mu, c}(\bar{p}_{c, \omega}, p) \overset{\scriptscriptstyle p \to 0}{\sim} (\omega Q^\textsc{a}_c/Q^\textsc{v}_c) \, \hat{\Gamma}^\textsc{v}_{\mu, c}(\bar{p}_{c, \omega}, p),
\end{equation} 
with $\bar{p}_{c, \omega} = (p_0, \omega \pfermi_c + p_1)$, must be imposed with a suitable choice of $Z^\textsc{a}_{\mu, c}$~\cite{Adler_2005}. Moreover, we add the counterterm $\sum_c \mu_c \int \psi^+_c \psi^-_c$ to the action~\eqref{eq:lattice_action} in order to control the shift of the Fermi points $\pm \pfermi_c$ due to the interaction. Our main result is the following.
\begin{theorem}
Suppose that in the lattice multiflavor Sommerfeld model~\eqref{eq:generating_functional} it is $a = M = 1$ and $e \le e_0$ for some small $e_0$, independent on $L,\beta$. If the Fermi points $\pfermi_c$ are all distinct and $\mu_c, Z^\textsc{a}_{\mu, c}$ are chosen as described above, $\hat{\Pi}^{\mu \nu}(p)$ exists in the limit $L, \beta \to +\infty$, it admits a convergent series expansion in $e$ and it satisfies
\begin{equation}
\label{eq:main_result}
ip_\alpha \hat{\Pi}^{\alpha \mu}(p) = -\frac{\epsilon^{\mu \alpha} p_\alpha}{\pi} \sum_c \frac{q_{+,c}^2 -q_{-,c}^2 }{4} + \mathcal{O}(\absto{p}{1 + \theta})
\end{equation}
for some constant $\theta \in (0, 1)$.
\end{theorem}
\paragraph{Sketch of the proof.}
After integrating out the massive field $A_\mu$ we obtain a purely fermionic theory with coupling $\lambda=e^2$ whose effective action contains field monomials of any degree. This theory can be analyzed with exact RG methods by writing $\psi^\pm_c = \sum_\omega \psi^\pm_{c, \omega} + \psi^\pm_{c, \textsc{uv}}$, where $\hat{\psi}^\pm_{c, \omega, k}$ is supported around $(0, \omega \pfermi_c)$ with a momentum cutoff; in addition, $\psi^\pm_{c, \omega}=\sum_{h=-\infty}^0
\psi^{\pm, h}_{c, \omega}$ with $\psi^{\pm, h}_{c, \omega}$
with a momentum cutoff selecting momenta $k$ in a shell around
$(0, \omega \pfermi_c)$ of size $2^h$.
After the integration of the fields 
$\psi^\pm_{c, \textsc{uv}}, \psi^{\pm, 0}_{c, \omega}, \dots ,\psi^{\pm, h+1}_{c, \omega}$ we get the following representation for the generating functional,
\begin{equation}
\label{eq:RG_representation}
W[J^\textsc{v}, J^\textsc{a},\phi]= \log \int \dgauss{g^{\le h}}{\psi} \, e^{-\mathscr{V}^h[\smash{\sqrt{Z_h}}\psi, J^\sharp, \phi]},
\end{equation}
holding for every negative integer $h$. The functional integral~\eqref{eq:RG_representation} depends on the propagator $g^{\le h}_{c, \omega}(\bar{k}_{c, \omega}) = \chi_h(k)[Z_{h, b}(-ik_0 + \omega \abs*{\mathfrak{v}_{h, b}}\! k_1)]^{-1} + \mathcal{O}(\absto{k}{\theta - 1})$, where the smooth function $\chi_h(k)$ selects momenta whose norms are less than $2^h$ (here and in the following, we let $b = (c, \omega)$). 
The effective potential $\mathscr{V}^h$ is a sum of monomials with $s$ $\psi$ fields and
$m$ $J$ fields, with kernels $W^h_{s, m}$. One can separate the relevant and marginal part
$\loc \mathscr{V}^h$ from the irrelevant part $\ren \mathscr{V}^h$ by writing $\mathscr{V}^h=\loc \mathscr{V}^h+ \ren \mathscr{V}^h$. The former is given by
\begin{multline}
\label{eq:loc_V}
\loc \mathscr{V}^h[\psi, J^\sharp, \phi] = \frac{1}{2}\sum_{bb'} \lambda_{h, bb'} \int_x \psi^+_{b, x} \psi^-_{b, x} \psi^+_{b', x} \psi^-_{b', x} \, + \\
+ \sum_b 2^h \mu_{h, b} \! \int_x \! \psi^+_{b, x} \psi^-_{b, x} - \sum_{b, \sharp, \mu} \frac{Z^\sharp_{h, \mu, b}}{Z_{h, b}} \int_x J^\sharp_{\mu, x} j^\sharp_{\mu, b, x}
\end{multline}
with $j^\textsc{v}_{\mu, b, x} = \psi^+_{b, x} \psi^-_{b, x}(1, i\omega)_\mu$ and $j^\textsc{a}_{\mu, b, x} = \omega j^\textsc{v}_{\mu, b, x}$. The latter contains the terms with $s/2+m>2$ or the non local quartic and quadratic terms. The kernels $W^h_{s, m}$ are expressed by convergent expansions in the running coupling constants
$\lambda_{k, bb'}$ with $k>h$, and the bound $\int \abs*{W^h_{s, m}} \lesssim C^{s+m} 2^{h(2 - s/2 - m)}$ holds if $\lambda_{k, bb'}$ is bounded. 

The bound for $W^h_{s, m}$ follows from Gram bounds for fermions and the tree expansion~\cite{Benfatto_Falco_Mastropietro_2010, Mastropietro_Porta_2022},
which ensure the convergence of the renormalized expansion and allow to establish non-perturbative regularity properties that will play a crucial role in the following.

The generic four-fermion monomial has the form $\int \psi^+_{c, \omega_1} \psi^-_{c, \omega_2} \psi^+_{c', \omega_3} \psi^-_{c', \omega_4}$, but momentum conservation requires that $(\omega_1 - \omega_2) \pfermi_c + (\omega_3 - \omega_4) \pfermi_{c'} = - (k_1 + k_2 - k_3 + k_4)_1$, where $k_1, \dots, k_4 = \mathcal{O}(2^h)$ are the momenta flowing inside $\psi^+_{c, \omega_1}, \dots, \psi^-_{c', \omega_4}$. If $\pfermi_c \ne \pfermi_{c'}$ and $h$ is sufficiently small, this condition is met \emph{only} provided that the four-fermion term has the structure displayed in~\eqref{eq:loc_V}. This is a major difference with respect to the Kogut-Susskind regularization. The couplings $\lambda_{h, bb'}$ converge to a line of fixed points $\lambda_{-\infty, bb'}$, which are analytic functions of $\lambda$. This is consequence of the asymptotic vanishing of the corresponding beta function~\cite{Benfatto_Falco_Mastropietro_2009, Mastropietro_Porta_2022}, which follows from an emergent chiral global $\mathrm{U}(1)$ symmetry (the latter would be broken if two different flavors shared the same Fermi momentum). Finally, the field strength renormalizations $Z_{h, b}$ and the current renormalizations $Z^\sharp_ {h, b}$ behave as $2^{\eta_b h}$ for some analytic $\eta_b = \eta_b(\lambda) = \mathcal{O}(\lambda^2)$. The counterterm $\mu_c$ can be chosen so that $\mu_{-\infty, c}=0$ and the parameters $t_c, \pfermi_c$ can be tuned so that $\abs*{\mathfrak{v}_{-\infty, b}}\!$ take arbitrary $\mathcal{O}(1)$ values (for instance, we may impose that $\abs*{\mathfrak{v}_{-\infty, b}}\! = 1$ to enforce emergent Lorentz invariance).

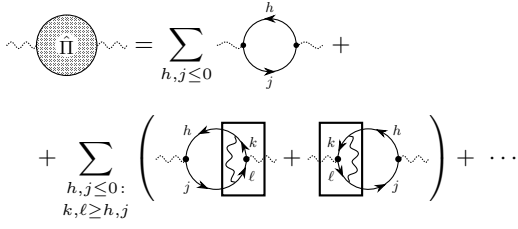
\begin{figure}[t]
\centering
\begin{equation}
\notag
\begin{multlined}
\begin{tikzpicture}[baseline={-0.6*height("$=$")}, scale=0.8]
\begin{feynman}
\draw[external photon]		(-1, 0)	--+(0.5, 0);
\draw[external photon]		(0.5, 0)	--+(0.5, 0);
\fill[draw=black, correlator background]		(0, 0)	circle	(0.5cm);
\node[circle, scale=0.8, fill=white, text=black,%
		inner sep=0, minimum size=1.2em]		(Pi)		at	(0, 0)	{$\hat{\Pi}$};
\end{feynman}
\end{tikzpicture}
= \sum_{h, j \le 0}
\begin{tikzpicture}[baseline={-0.6*height("$=$")}, scale=0.7]
\begin{feynman}
\draw[external photon]	(-1, 0)	--+(0.5, 0);
\draw[external photon]	(0.5, 0)	--+(0.5, 0);
\draw[with arrow]	(0.5, 0)		arc	(0:180:0.5);
\draw[with arrow]	(-0.5, 0)	arc	(180:360:0.5);
\node[scale=0.6]		(h)			at	($1.43*(90:0.5cm)$)		{$h$};
\node[scale=0.6]		(j)			at	($1.43*(-90:0.5cm)$)		{$j$};
\fill[black]	(-0.5, 0)	circle	(1.5pt);
\fill[black]	(0.5, 0)	circle	(1.5pt);
\end{feynman}
\end{tikzpicture} \,\, + \\[4pt]
+ \sum_{\substack{h, j \le 0 \colon\!\! \\[2pt] k, \ell \ge h, j}}
\left(
\begin{tikzpicture}[baseline={-0.6*height("$=$")}, scale=0.8]
\begin{feynman}
\draw[external photon]	(-1, 0)	--+(0.5, 0);
\draw[external photon]	(0.5, 0)	--+(0.5, 0);
\draw[photon]			(50:0.5cm)	to [out=240, in=120]	(-50:0.5cm);
\draw[with arrow]	(0.5, 0)			arc	(0:50:0.5);
\draw[with arrow]	(50:0.5cm)		arc	(50:180:0.5);
\draw[with arrow]	(-0.5, 0)		arc	(180:310:0.5);
\draw[with arrow]	(310:0.5cm)		arc	(310:360:0.5);
\draw[black, thick]	(0.1, -0.6)		rectangle	(0.8, 0.6);
\node[scale=0.6]		(a)	at	($1.35*(135:0.5cm)$)	{$h$};
\node[scale=0.6]		(b)	at	($1.35*(225:0.5cm)$)	{$j$};
\node[scale=0.6]		(c)	at	($1.35*(25:0.5cm)$)	{$k$};
\node[scale=0.6]		(d)	at	($1.3*(-25:0.5cm)$)	{$\ell$};
\fill[black]	(-0.5, 0)	circle	(1.5pt);
\fill[black]	(0.5, 0)	circle	(1.5pt);
\end{feynman}
\end{tikzpicture}
+
\begin{tikzpicture}[baseline={-0.6*height("$=$")}, scale=0.8]
\begin{feynman}
\draw[external photon]	(-1, 0)	--+(0.5, 0);
\draw[external photon]	(0.5, 0)	--+(0.5, 0);
\draw[photon]		(-130:0.5cm)	to [out=60, in=300]	(130:0.5cm);
\draw[with arrow]	(0.5, 0)			arc	(0:130:0.5);
\draw[with arrow]	(130:0.5cm)		arc	(130:180:0.5);
\draw[with arrow]	(-0.5, 0)		arc	(180:230:0.5);
\draw[with arrow]	(230:0.5cm)		arc	(230:360:0.5);
\draw[black, thick]	(-0.8, -0.6)		rectangle	(-0.1, 0.6);
\node[scale=0.6]		(a)	at	($1.35*(45:0.5cm)$)	{$h$};
\node[scale=0.6]		(b)	at	($1.3*(-45:0.5cm)$)	{$j$};
\node[scale=0.6]		(c)	at	($1.35*(155:0.5cm)$)	{$k$};
\node[scale=0.6]		(d)	at	($1.35*(205:0.5cm)$)	{$\ell$};
\fill[black]	(-0.5, 0)	circle	(1.5pt);
\fill[black]	(0.5, 0)	circle	(1.5pt);
\end{feynman}
\end{tikzpicture}
\right)
+ \,\, \cdots
\end{multlined}
\end{equation}
\caption{The first few terms of the multiscale expansion of $\hat{\Pi}^{\mu \nu}(p)$. The variables $h, k, j, \ell$ are the scale labels of the propagators. Black rectangles are used to evidence subdiagrams that need to be \emph{renormalized}, that is, deprived of their local parts. Furthermore, black dots are used to denote the presence of $Z_{s, b}^\sharp$ factors; fermion lines carry $1/Z_{s, b}$ factors.}
\label{fig:first_few_diagrams}
\end{figure}

The outcome of the RG analysis is a renormalized convergent expansion for the correlation  $\hat{\Pi}^{\mu \nu}(p)$; an example of the first terms is shown in Fig~\ref{fig:first_few_diagrams}. In the non-interacting case only the first term,
the bubble graph, is present. In presence of interaction two things happen; the first is that the bubble acquires renormalizations, both in the wave function and in the vertices; the second is that 
there are infinitely many other diagrams contributing to the anomaly.
The validity of the universality property~\eqref{eq:main_result} relies on a huge cancellation between graphs which is impossible to see directly from the expansion.

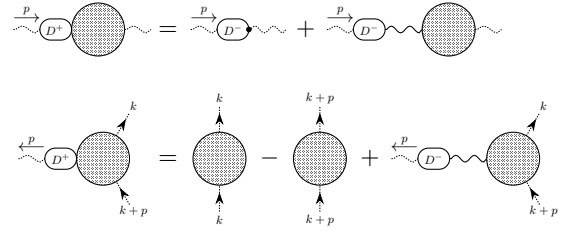
\begin{figure}[t]
\begin{align*}
\begin{tikzpicture}[baseline={-0.6*height("$=$")}, scale=0.7]
\begin{feynman}
\draw[external photon]	(-1.6, 0)	--+(0.5, 0);
\draw[external photon]	(0.5, 0)		--+(0.5, 0);
\draw[rounded corners]	(-0.5, -0.2)	rectangle	(-1.1, 0.2);
\fill[draw, correlator background]		(0, 0)	circle	(0.5cm);
\node[scale=0.5]		(p)			at	(-1.35, 0.35)	{$p$};
\node[scale=0.7]		(arr)		at	(-1.35, 0.2)		{$\longrightarrow$};
\node[scale=0.5]		(D)			at	(-0.82, 0)		{$D^+$};
\end{feynman}
\end{tikzpicture}
& =
\begin{tikzpicture}[baseline={-0.6*height("$=$")}, scale=0.7]
\begin{feynman}
\draw[external photon]			(-0.2, 0)		--	(0.5, 0);
\draw[external photon]			(-1.3, 0)	--	(-0.8, 0);
\fill[black]						(-0.2, 0)	circle	[radius=1.5pt];
\draw[rounded corners]			(-0.2, -0.2)	rectangle	(-0.8, 0.2);	
\node[scale=0.5]		(D)			at	(-0.5, 0)		{$D^-$};
\node[scale=0.5]		(p)			at	(-1.05, 0.35)	{$p$};
\node[scale=0.7]		(arr)		at	(-1.05, 0.2)		{$\longrightarrow$};
\end{feynman}
\end{tikzpicture} \, +
\begin{tikzpicture}[baseline={-0.6*height("$=$")}, scale=0.7]
\begin{feynman}
\draw[external photon]	(-2.3, 0)	--+(0.5, 0);
\draw[external photon]	(0.5, 0)		--+(0.5, 0);
\draw[photon]			(-1.2, 0)	--	(-0.5, 0);
\draw[rounded corners]	(-1.2, -0.2)	rectangle	(-1.8, 0.2);	
\fill[draw, correlator background]		(0, 0)	circle	(0.5cm);
\node[scale=0.5]		(p)			at	(-2.05, 0.35)	{$p$};
\node[scale=0.7]		(arr)		at	(-2.05, 0.2)		{$\longrightarrow$};
\node[scale=0.5]		(D)			at	(-1.5, 0)		{$D^-$};
\end{feynman}
\end{tikzpicture} \\[8pt]
\begin{tikzpicture}[baseline={-0.6*height("$=$")}, scale=0.7]
\begin{feynman}
\draw[external photon]	(-1.6, 0)	-- (-1.1, 0);
\draw[rounded corners]	(-1.1, -0.2)		rectangle	(-0.5, 0.2);
\draw[external fermion]	(60:0.5cm)	--	(60:1cm);
\draw[external fermion]	(-60:1cm)	--	(-60:0.5cm);
\fill[draw, correlator background]	(0, 0)	circle	(0.5cm);
\node[scale=0.5]		(D)		at	(-0.8, 0)	{$D^+$};
\node[scale=0.5]		(p)			at	(-1.35, 0.35)	{$p$};
\node[scale=0.7]		(arr)		at	(-1.35, 0.2)		{$\longleftarrow$};
\node[scale=0.5]		(k)			at	(60:1.15cm)		{$k$};
\node[scale=0.5]		(kp)			at	(-60:1.15cm)		{$k+p$};
\end{feynman}
\end{tikzpicture}
& = \,\,
\begin{tikzpicture}[baseline={-0.6*height("$=$")}, scale=0.7]
\begin{feynman}
\draw[external fermion]				(0, -1)	--	(0, -0.5);
\draw[external fermion]				(0, 0.5)	--	(0, 1);
\fill[draw, correlator background]	(0, 0)	circle	(0.5cm);
\node[scale=0.5]		(k1)		at	(0, -1.15)		{$k$};
\node[scale=0.5]		(k2)		at	(0, 1.15)		{$k$};
\end{feynman}
\end{tikzpicture}
\,\, - \,\,
\begin{tikzpicture}[baseline={-0.6*height("$=$")}, scale=0.7]
\begin{feynman}
\draw[external fermion]				(0, -1)	--	(0, -0.5);
\draw[external fermion]				(0, 0.5)	--	(0, 1);
\fill[draw, correlator background]	(0, 0)	circle	(0.5cm);
\node[scale=0.5]		(kp1)	at	(0, -1.15)	{$k+p$};
\node[scale=0.5]		(kp2)	at	(0, 1.15)	{$k+p$};
\end{feynman}
\end{tikzpicture}
\,\, + 
\begin{tikzpicture}[baseline={-0.6*height("$=$")}, scale=0.7]
\begin{feynman}
\draw[external photon]	(-2.3, 0)	-- 	(-1.8, 0);
\draw[photon]			(-1.2, 0)	-- 	(-0.5, 0);
\draw[rounded corners]	(-1.8, -0.2)		rectangle	(-1.2, 0.2);
\draw[external fermion]	(60:0.5cm)	--	(60:1cm);
\draw[external fermion]	(-60:1cm)	--	(-60:0.5cm);
\fill[draw, correlator background]	(0, 0)	circle	(0.5cm);
\node[scale=0.5]		(D)		at	(-1.5, 0)	{$D^-$};
\node[scale=0.5]		(p)			at	(-2.05, 0.35)	{$p$};
\node[scale=0.7]		(arr)		at	(-2.05, 0.2)		{$\longleftarrow$};
\node[scale=0.5]		(k)			at	(60:1.15cm)		{$k$};
\node[scale=0.5]		(kp)			at	(-60:1.15cm)		{$k+p$};
\end{feynman}
\end{tikzpicture}
\end{align*}
\caption{Anomalous Ward Identities for the correlators $\expval*{\rho_b, \rho_{b'}}_\rff$ and $\expval*{\rho_b, \psi^-_{b'}, \psi^+_{b'}}_\rff$. The anomalous terms consist in the whole right hand side of the first line and in the last term of the second line.}
\label{fig:anomalous_WI}
\end{figure}
The idea is to introduce another regularization of the multiflavor Sommerfield model, which we call \emph{reference model}, defined in the continuum with a momentum regularization.
It is expressed in terms of Weyl fermions and is regularized with a momentum cutoff. The fermion propagator is
\begin{equation}
\hat{g}^\rff_b(k) = \chi_N(\abs*{D_b(k)}\!) [\mathcal{Z}_b D_b(k)]^{-1},
\end{equation}
where $\chi_N$ is a smooth cutoff function that is equal to $1$ on $[0, 2^N]$
and $D^\pm_b(k) \dff -ik_0 \pm \mathfrak{u}_b k_1$. The partition function of the reference model is $\cramped{\int \dgauss{g^\rff}{\psi} \, e^{-V^\rff[\cramped{\sqrt{\mathcal{Z}}}\psi]}}$, where $V^\rff[\psi]$ is an interaction between the bilinears $\rho_b \equiv \rho_{c, \omega} = \psi^+_{c, \omega} \psi^-_{c, \omega}$ mediated by a short\-/ranged potential $v$ such that $\hat{v}_0 = 1$, namely
\begin{equation}
\label{eq:reference_model_action}
V^\rff[\psi] = \frac{1}{2} \sum_{b b'} \alpha_{b b'} \! \int \dd{x} \dd{y} \, \rho_b(x) \rho_{b'}(y) \, v(x-y)
\end{equation}
with $\alpha_{b b'} \dff \lambda^\rff_{bb'} \mathcal{Z}_b \mathcal{Z}_{b'}$. 
The vector current and the axial current are respectively given by $j_\mu^\sharp = \sum_b \rho_b \, \cramped{\sigma_{\mu, b}^\sharp}$, with $\sigma_{\mu, b}^\textsc{v} = (\mathcal{Z}^\textsc{v}_{0, c, \omega}, i\omega \mathcal{Z}^\textsc{v}_{1, c, \omega})_\mu$ and $\sigma_{\mu, b}^\textsc{a} = (\omega \mathcal{Z}^\textsc{a}_{0, c, \omega}, i \mathcal{Z}^\textsc{a}_{1, c, \omega})_\mu$. To obtain insertions of $j^\sharp_\mu$ inside a correlation function, we introduce the source term $\sum_\sharp \int J^\sharp \cdot j^\sharp$ and we differentiate the generating functional with respect to $J^\sharp_\mu$. The correlation functions of the reference model are $\hat{S}^\rff_b(k) = \expval*{\hat{\psi}^-_{b, k}; \hat{\psi}^+_{b, k}}_\rff$, $\hat{A}_{bb'}^\rff(p) = \expval*{\hat{\rho}_{b, p}; \hat{\rho}_{b', -p}}_\rff$ and $\hat{\Gamma}_{bb'}^\rff(k, p) = \expval*{\hat{\rho}_{b, p}; \hat{\psi}^-_{b', k}; \hat{\psi}^+_{b', k+p}}_\rff$.

The exact RG analys allows to make precise the relations between the correlations of the model~\eqref{eq:lattice_action} and those of the reference model. It is indeed possible to fine-tune the parameters 
$\mathfrak{u}_b, \mathcal{Z}_b, \lambda^\rff_{bb'}$ of the reference model so that~\cite{Mastropietro_Porta_2022, Benfatto_Falco_Mastropietro_2010}
\begin{equation}
\label{eq:Pi_decomposition}
\hat{\Pi}^{\mu \nu}(p) = \hat{\Pi}^{\mu \nu}_\rff(p) + \mathscr{R}^{\mu \nu}(p).
\end{equation}
Formula~\eqref{eq:Pi_decomposition} says that the correlation of the lattice model is equal to the correlation of the reference model with fine-tuned parameters, up to a term
$\mathscr{R}^{\mu \nu}(p)$ which is more regular as a function of $p$; in particular, $\mathscr{R}^{\mu \nu}(p)$ is H\"{o}lder-continuous while
$\hat{\Pi}^{\mu \nu}_\rff(p)$ is not continuous. Similarly, $\hat{S}_c(\bar{k}_{c, \omega}) = \hat{S}^\rff_b(k) + \mathcal{O}(\absto{k}{\theta - 1})$, $\hat{\Gamma}^{\sharp}_{\mu, c}(\bar{p}_{c, \omega}, p) = \hat{\Gamma}^{\textsc{r}, \sharp}_{\mu, b}(p, p)+\mathcal{O}(\absto{p}{\theta - 2})$.
The proof of (\ref{eq:Pi_decomposition}) relies on the convergence of the expansion combined with the choice of a momentum regularization for the reference model, which is the natural one 
to make contact with the RG analysis of the lattice model. 

Observe that the reference model is not solvable by bosonization. Nevertheless, exact nonperturbative expressions for the correlation functions can be obtained thanks to the validity of Ward Identities~\cite{Benfatto_Falco_Mastropietro_2010, Mastropietro_Porta_2022}: letting $\Delta \hat{S}_b^\rff(k, p) \dff \hat{S}^\rff_b(k) - \hat{S}^\rff_b(k+p)$, we have
\begingroup
\allowdisplaybreaks
\begin{align}
\label{eq:anomalous_WI_Gamma}
D^+_{b}(p) \sum_{b''}(T_{b b''}^{-1})(p) \,
\mathcal{Z}_{b''} \hat\Gamma_{b'' b'}^\rff(k, p)
& = \delta_{bb'} \Delta \hat{S}^\rff_b(k, p) \\
\label{eq:anomalous_WI_rho}
D^+_{b}(p) \sum_{b''} (T_{b b''}^{-1})(p) \,
\mathcal{Z}_{b''} \hat{A}_{b'' b'}^\rff(p) 
& = -\frac{D^-_b(p)}{4\pi \abs*{\mathfrak{u}_b} \!} \frac{\delta_{bb'}}{\mathcal{Z}_{b}}
\end{align}
\endgroup
with $\cramped{(T_{bb'}^{-1})(p)} = \delta_{bb'} - \cramped{[4\pi \abs*{\mathfrak{u}_b} \! D_b^+(p)]^{-1}\lambda_{bb'}^\rff D_b^-(p)}$ and $D^\pm_b(p) = -ik_0 \pm \mathfrak{u}_b k_1$. As a consequence, we can write
\begin{equation}
\label{eq:Pi_explicit}
\hat{\Pi}^{\mu \nu}_\rff = - \sum_{bb'} \sigma^\textsc{a}_{\mu, b} \sigma^\textsc{v}_{\nu, b'} T_{bb'} D^-_{b'} (4\pi \abs*{\mathfrak{u}_{b'} \!} \mathcal{Z}_b \mathcal{Z}_{b'} D^+_{b'})^{-1}.
\end{equation}
Formula~\eqref{eq:Pi_explicit} is obtained by solving the anomalous Ward Identity~\eqref{eq:anomalous_WI_rho} for $\hat{A}_{bb'}^\rff(p)$ and subsequently exploiting the fact that $\hat{\Pi}^{\mu \nu}_\rff(p) = \sum_{bb'} \sigma^\textsc{a}_{\mu, b} \sigma^\textsc{v}_{\nu, b'} \hat{A}^\rff_{bb'}(p)$.
In the reference model there is invariance under global and chiral $\mathrm{U}(1)$ phase symmetry and this allows to get an explicit expression for the current correlations; on the other hand, the WI are anomalous and the vector current is not conserved, in contrast to what happens in the lattice mode where the vector current is conserved~\eqref{eq:lattice_WI_current}.

In conclusion, we get
\begin{equation}
\label{eq:Pi_reference_model_comparison}
\hat{\Pi}^{\mu \nu}(p) = - \sum_{bb'} \frac{ \sigma^\textsc{a}_{\mu, b} \sigma^\textsc{v}_{\nu, b'}T_{bb'} D^-_{b'}}{4\pi \abs*{\mathfrak{u}_{b'} \!} \mathcal{Z}_b \mathcal{Z}_{b'} D^+_{b'}}+ \mathscr{R}^{\mu \nu}(0) + \mathcal{O}(\absto{p}{\theta})
\end{equation}
where we have used the H\"{o}lder continuity of $\mathscr{R}^{\mu \nu}(p)$ to write 
$\mathscr{R}^{\mu \nu}(p) = \mathscr{R}^{\mu \nu}(0) + \mathcal{O}(\absto{p}{\theta})$. We have therefore an expression depending on 
$2n^2 + 11n$ parameters $(\lambda^\rff_{bb'}, \mathcal{Z}_b, \mathcal{Z}_{\mu, b}^\textsc{v}, \mathcal{Z}_{\mu, b}^\textsc{a}, \mathfrak{u}_b)$ and 4 further unknowns ($\mathscr{R}^{\mu \nu}(0)$), all expressed by series expansions depending on all the lattice irrelevant terms.

These parameters are however not independent and we can see whether the Ward identities of the lattice model and of the reference model give us sufficiently many constraints to fix them.
We first note that the vertex functions of the lattice model and those of the reference model are the same up to subleading terms, as we have chosen the bare parameters of the reference model just to enforce this. Such condition provides relations between the parameters. Indeed, the simultaneous validity of lattice~\eqref{eq:lattice_WI_vertex} and chiral~\eqref{eq:anomalous_WI_Gamma} Ward Identities for the vertex function implies the following remarkable relations:
\begin{equation}
\label{eq:constraints}
\epsilon_{\omega'}^\sharp Q^\sharp_{c'}  + \sum_{\mu,b} ip_\mu \sigma^\sharp_{\mu,b}\frac{T_{bb'}(p)}{ \mathcal{Z}_b D^+_{b'}(p)} = 0
\end{equation}
which are valid up to $\mathcal{O}(\absto{p}{\theta})$ errors; here, $\epsilon^\textsc{a}_{\omega'} = \omega'$ and $\epsilon^\textsc{v}_{\omega'} = 1$. As a consequence of~\eqref{eq:constraints} and~\eqref{eq:Pi_explicit}, we obtain
\begin{equation}
\label{eq:contractions}
\begin{aligned}[c]
ip_\alpha \hat{\Pi}^{\alpha \mu}_\rff(p) 
& = -p^\alpha \sum_b (i, \omega \abs*{\mathfrak{u}_b} \!)_\alpha \frac{\sigma^\textsc{v}_{\mu, b} \, \omega Q^\textsc{a}_c}{4\pi \abs*{\mathfrak{u}_b} \! \mathcal{Z}_b} \\
ip_\alpha \hat{\Pi}^{\mu \alpha}_\rff(p) 
& = -p^\alpha \sum_b (i, \omega \abs*{\mathfrak{u}_b} \!)_\alpha \frac{\sigma^\textsc{a}_{\mu, b} Q^\textsc{v}_c}{4\pi \abs*{\mathfrak{u}_b} \! \mathcal{Z}_b}
\end{aligned}
\end{equation}
again up to $\mathcal{O}(\absto{p}{1 + \theta})$ errors. 
We use the property that $\mathscr{R}^{\mu \nu}(p)$ is continuous
to write $ip_\mu \hat{\Pi}^{\mu \nu}(p) = ip_\mu \hat{\Pi}_\rff^{\mu \nu}(p) + i p_\mu \mathscr{R}^{\mu \nu}(0) + \mathcal{O}(\absto{p}{1 + \theta})$, where the first term is given by the first of~\eqref{eq:contractions}. The WI~\eqref{eq:lattice_WI_current} fixes however also the value of the contribution of the second term; it is sufficient to notice that $i p_\nu [\hat{\Pi}^{\mu \nu}_\rff(p) + \mathscr{R}^{\mu \nu}(0)] = 0$ up to h.\ o.\ t. in $p$ to get
\begin{equation}
\label{eq:R_constant_determination}
i\mathscr{R}^{\mu \nu}(0) = -\partial [i p_\alpha \hat{\Pi}^{\mu \alpha}_\rff(p)]/\partial p_\nu \vert_0.
\end{equation}
After combining the above relations with~\eqref{eq:contractions}, we finally obtain $ip_\alpha \hat{\Pi}^{\alpha \mu}(p) = \sum_b M^{\mu \alpha}_b p_\alpha$ up to subdominant corrections, where the matrix $M_b$ is equal to
\begin{equation}
\label{eq:M_matrix}
\frac{1}{4\pi\mathcal{Z}_b}
\begin{pmatrix}
\displaystyle i\omega\frac{\mathcal{Z}^\textsc{a}_{0, b} Q^\textsc{v}_c - \mathcal{Z}^\textsc{v}_{0, b} Q^\textsc{a}_c}{\abs*{\mathfrak{u}_b}\!} 
&  & -\mathcal{Z}^\textsc{v}_{0, b} Q^\textsc{a}_c - \displaystyle \frac{\mathcal{Z}^\textsc{a}_{1, b}Q^\textsc{v}_c}{\abs*{\mathfrak{u}_b}\!} \\[10pt]
\displaystyle \frac{\mathcal{Z}^\textsc{v}_{1, b} Q^\textsc{a}_c}{\abs*{\mathfrak{u}_b}\!} + \mathcal{Z}^\textsc{a}_{0, b} Q^\textsc{v}_c
&  & i\omega(\mathcal{Z}^\textsc{a}_{1, b} Q^\textsc{v}_c - \mathcal{Z}^\textsc{v}_{1, b} Q^\textsc{a}_c)
\end{pmatrix}.
\end{equation}
The entries of $M_b$ apparently depend on the bare parameters of the reference model; moreover, it is not obvious \emph{a priori} that $M_b$ depends on $Q^\sharp_c$ only through the combination $\sum_c Q^\textsc{v}_c Q^\textsc{a}_c$, as stated in the main result. To solve these problems, we exploit once again the constraint~\eqref{eq:constraints}, which can be converted into a linear system that can be solved for $\mathcal{Z}^\sharp_{\mu, b}/\mathcal{Z}_b$. The result is
\begin{align}
\label{eq:Z0_determination}
\mathcal{Z}^\sharp_{0, b}/\mathcal{Z}_b
& = Q^\sharp_c - \sum_{b'} Q^\sharp_{c'} \epsilon^\sharp_\omega \epsilon^\sharp_{\omega'} \lambda^\rff_{b'b}(4\pi\abs*{\mathfrak{u}_{b'} \!}\!)^{-1} \\
\label{eq:Z1_determination}
\mathcal{Z}^\sharp_{1, b}/\mathcal{Z}_b
& = Q^\sharp_c \, \abs*{\mathfrak{u}_b} + \sum_{b'} Q^\sharp_{c'} \, \omega \omega' \epsilon^\sharp_\omega \epsilon^\sharp_{\omega'} \lambda_{b'b}^\rff(4\pi)^{-1}
\end{align}
In order to prove~\eqref{eq:Z0_determination} and~\eqref{eq:Z1_determination}, we note that although $T_{bb'}(p)$ has no limit as $p \to 0$, the limits $T_{bb'}^- = \lim_{p_0 \to 0} \allowbreak \lim_{p_1 \to 0} T_{bb'}(p)$ and $T_{bb'}^+ = \lim_{p_1 \to 0} \allowbreak \lim_{p_0 \to 0} T_{bb'}(p)$ do exist; in particular, $(T^\pm)^{-1}_{bb'} = \delta_{bb'} \pm (4\pi\abs*{\mathfrak{u}_b}\!)^{-1} \lambda^\rff_{bb'}$. As a consequence, if we take the $\lim_{p_0 \to 0} \allowbreak \lim_{p_1 \to 0}$ limit of both sides of~\eqref{eq:constraints} with $\sharp = \textsc{v}$, we obtain the linear system $Q_{c'}^\textsc{v} = \sum_b (\mathcal{Z}^\textsc{v}_{0, b}/\mathcal{Z}_b) T^-_{bb'}$, which is solved by $\mathcal{Z}^\textsc{v}_{0, b}/\mathcal{Z}_b = Q^\textsc{v}_c - \sum_{b'} Q^\textsc{v}_{c'} \lambda^\rff_{b'b}(4\pi \abs*{\mathfrak{u}_{b'}}\!)^{-1}$. If we consider the same limit with $\sharp = \textsc{a}$, we obtain the linear system $\omega' Q_{c'}^\textsc{a} = \sum_b (\omega \mathcal{Z}^\textsc{a}_{0, b}/\mathcal{Z}_b) T^-_{bb'}$, whose solution is instead $\omega \mathcal{Z}^\textsc{a}_{0, b}/\mathcal{Z}_b = \omega Q^\textsc{a}_c - \sum_{b'} \omega' Q^\textsc{a}_{c'} \lambda^\rff_{b'b} (4\pi\abs*{\mathfrak{u}_{b'}}\!)^{-1}$. The two relations found in this way coincide with~\eqref{eq:Z0_determination}. Formulae~\eqref{eq:Z1_determination} can be proved with an analogous procedure by taking the $\lim_{p_1 \to 0} \lim_{p_0 \to 0}$ limit of~\eqref{eq:constraints}. 

By plugging~\eqref{eq:Z0_determination} and~\eqref{eq:Z1_determination} into~\eqref{eq:M_matrix}, we can prove the relation $\sum_b M_b^{\alpha \mu} = -(\epsilon^{\alpha \mu}/\pi) \sum_c Q^\textsc{v}_c Q^\textsc{a}_c$, which readily implies our main result. We start from the entry $\sum_b M^{01}_b$, which becomes
\begin{multline}
\label{eq:M01_example}
-\sum_c \frac{Q^\textsc{v}_c Q^\textsc{a}_c}{\pi} + \frac{1}{4\pi} \sum_{bb'} \lambda^\rff_{b'b} \! \left[ \frac{Q^\textsc{v}_{c'} Q^\textsc{a}_c}{\abs*{\mathfrak{u}_{b'}}\!} - \frac{Q^\textsc{v}_c Q^\textsc{a}_{c'}}{\abs*{\mathfrak{u}_b}\!} \right]
\end{multline}
after using~\eqref{eq:Z0_determination} and~\eqref{eq:Z1_determination} to eliminate $\mathcal{Z}^\textsc{v}_{0, b}/\mathcal{Z}_b$ and $\mathcal{Z}^\textsc{a}_{1, b}/\mathcal{Z}_b$ inside~\eqref{eq:M_matrix}. Since the term in square brackets is antisymmetric under the exchange $b \leftrightarrow b'$ and $\lambda^\rff_{bb'} = \lambda^\rff_{b'b}$, the whole second addend of~\eqref{eq:M01_example} vanishes, so $\sum_b M^{01} = -\sum_c Q^\textsc{v}_c Q^\textsc{a}_c/\pi$. Similarly, the entry $\sum_b M^{10}_b$ is equal to
\begin{equation}
\sum_c \frac{Q^\textsc{v}_c Q^\textsc{a}_c}{\pi} + \frac{1}{4\pi} \sum_{bb'} \lambda^\rff_{b'b} \! \left[ \frac{Q^\textsc{v}_{c'} Q^\textsc{a}_c}{\abs*{\mathfrak{u}_b}\!} - \frac{Q^\textsc{v}_c Q^\textsc{a}_{c'}}{\abs*{\mathfrak{u}_{b'}}\!} \right] \omega \omega'
\end{equation}
and the second term vanishes as well, so $\sum_b M^{10} = \sum_c Q^\textsc{v}_c Q^\textsc{a}_c/\pi$. Analogous cancellations occur with the other entries, leading us to $\sum_b M^{00} = \sum_b M^{11} = 0$. All these cancellations are tightly related with the \emph{discontinuity} of the matrix $T_{bb'}(p)$ at $p = 0$. 

\paragraph{Conclusions.}
We have provided the first rigorous nonperturbative proof of chiral anomaly cancellation in an interacting renormalizable two-dimensional lattice quantum field theory and established the universality of the anomaly-free condition. Rather than being mere corrections to scaling, lattice irrelevant operators are shown to provide essential contributions to the cancellation mechanism.
Our result identifies a new mechanism by which continuum consistency conditions emerge from interacting lattice theories, which we
expect to be of general applicability; natural extensions to more realistic models include the case where the external chiral field $Z_\mu$ is promoted to a quantum field and, ultimately, to four-dimensional theories.

\bibliography{Bibliography_PRL}

\clearpage

\onecolumngrid

\newgeometry{left = 3.2cm, right = 3.2cm, top = 3.2cm, bottom = 3.2cm}

\section{Supplementary material}
\label{sec:supplementary}

\subsection{I - Lattice anomaly in the non-interacting case}
\label{suppl:renormalized_expansion}
We specialize our strategy to compute the anomaly to the non-interacting case, where calculations are straightforward. In the absence of interactions ($e = 0$), the axial-vector correlation is given by a single bubble graph plus a \emph{Schwinger term} $K_{\mu \nu}$, namely
\begin{equation}
\label{eq:Pi_Wick}
\begin{aligned}[c]
\hat{\Pi}^{\mu \nu}(p) 
& = i\sum_{cc', \alpha} Z^\textsc{a}_{\mu, c} Q^\textsc{a}_c Q^\textsc{v}_{c'} \epsilon_{\mu \alpha} \frac{\delta_{cc'}}{L\beta} \sum_q u_{c, \alpha}(q) u_{c', \nu}(p + q) \hat{g}_c(q) \hat{g}_{c'}(p+q) + K_{\mu \nu} \\
& =
\begin{tikzpicture}[baseline={-0.6*height("$=$")}, scale=0.8]
\begin{feynman}
\draw[external photon]	(-1, 0)	--+(0.5, 0);
\draw[external photon]	(0.5, 0)	--+(0.5, 0);
\draw[with arrow]	(0.5, 0)		arc	(0:180:0.5);
\draw[with arrow]	(-0.5, 0)	arc	(180:360:0.5);
\end{feynman}
\end{tikzpicture}
\,\,
+
\,\,
\begin{tikzpicture}[baseline={-0.6*height("$=$")}, scale=0.8]
\begin{feynman}
\fill[draw=black, correlator background]	(0, 0)	circle	(0.5cm);
\node[circle, scale=0.8, fill=white, text=black,%
		inner sep=0, minimum size=1.2em]		(K)		at	(0, 0)	{$K$};
\end{feynman}
\end{tikzpicture}
\end{aligned}
\end{equation}
where $u_{c, \mu}(q) = (1, it_c\sin(aq_1))_\mu$. The Schwinger term arises because the axial current $j^\textsc{a}_\mu$ is not a local operator; its explicit form is
\begin{equation}
K_{\mu \nu} = i a \, \delta_{\mu, 0} \, \delta_{\nu, 1} \sum_c Q^\textsc{a}_c Q^\textsc{v}_c t_c Z^\textsc{a}_{1, c} \expval*{\psi^+_{c, 0} e^{-iae A_{1, 0}} \psi^-_{c, ae_1} - \psi^+_{c, ae_1} e^{iae A_{1, 0}} \psi^-_{c, 0}}.
\end{equation} 
The axial charge normalization condition requires to fix $Z^\textsc{a}_{0, c} = 1/\abs*{\mathfrak{v}_c}$ and $Z^\textsc{a}_{1, c} = \abs*{\mathfrak{v}_c}\!$ in this non-interacting case.
The propagator is $\hat g_{c}(k) = (-ik_0 - t_c[\cos(ak_1) - \cos(a \pfermi_c)]/a)^{-1}$ and it is singular at $(0, \pm \pfermi_c)$.

One could explicitly evaluate the integrals in the above expression, but we follow instead the strategy outlined in the main text, specialized to the non interacting case.
Setting $1 \equiv \sum_{\omega=\pm} \chi(q-(0, \omega \pfermi_c)) + \bar{\chi}(q)$ where $\chi(q-(0, \omega \pfermi_c))$
is a smooth function with compact support around the points $(0, \omega \pfermi_c)$, we write
\begin{equation}
\label{eq:propagator_poles_splitting}
\hat{g}_c(q)=\sum_{\omega=\pm} \chi(q-(0, \omega \pfermi_c)) \hat{g}_c(q) + \bar{\chi}(q)\hat{g}_c(q) \equiv \sum_{\omega = \pm} \hat{g}_{c, \omega}(q_0, \omega \pfermi_c + q_1) + \bar{\chi}(q)\hat{g}_c(q).
\end{equation}
with $\abs*{\bar{\chi}(q)\hat{g}_c(q)}$ bounded. 
Setting $k=q-(0, \omega \pfermi_c)$
we further decompose $\hat{g}_{c, \omega}(q_0, \omega \pfermi_c + q_1)$ in a linear relativistic propagator $\chi(k)/D_\omega(k)$, singular at $k=0$, and a remainder which is also bounded. In this way we have written the lattice propagator as sum of two chiral relativistic propagators and a subleading correction. Correspondingly, the correlation can be written as
\begin{equation}
\label{eq:Pi_loop}
\hat{\Pi}^{\mu \nu}(p) = - \sum_{c, \omega} Q^\textsc{a}_c Q^\textsc{v}_c 
\begin{pmatrix}
	\omega
&	i \abs*{\mathfrak{v}_c}  \\
	i \abs*{\mathfrak{v}_c}
&	-\omega \absto{\mathfrak{v}_c}{2} 	\\
\end{pmatrix}_{\! \mu \nu}
\int \frac{\dd[2]{q}}{(2\pi)^2} \frac{\chi_{c, \omega}(q) \chi_{c, \omega}(p + q)}{D^+_{c, \omega}(q) D^+_{c, \omega}(p+q)} + \mathscr{R}^{\mu \nu}(p)
\end{equation}
in the $\beta, L \to +\infty$ limit. The function $\mathscr{R}^{\mu \nu}(p)$ accounts for the constant $K_{\mu \nu}$ and for the contributions coming from the $\bar{\chi}(q)\hat{g}_c(q)$ term displayed in~\eqref{eq:propagator_poles_splitting}, the $\hat{r}_c(q)$ term displayed in~\eqref{eq:propagator_linearization} and the Taylor remainders of the expansion of $u_{c, \alpha}(q)$ around $(0, \pm \pfermi_c)$. 
Note that $\mathscr{R}^{\mu \nu}(p)$ is H\"{o}lder-continous 
while the first integral is not continuous at $p=0$. Formula~\eqref{eq:Pi_loop} is the analogous of~\eqref{eq:Pi_decomposition} in the non-interacting case; the first term is the contribution from Weyl fermions with linear dispersion relation with momentum regularization (the reference model defined by~\eqref{eq:reference_model_action} in the $\lambda_{bb'}^\rff=0$ case) and the second term comes from the non linear lattice corrections. 

By H\"{o}lder-continuity  we can write
$$\mathscr{R}^{\mu \nu}(p) = \mathscr{R}^{\mu \nu}(0) + \mathcal{O}(a^\theta \absto{p}{\theta})$$
for some $\theta > 0$ (indeed, $\theta=1$ in this case).
We get therefore
\begin{equation}
\label{eq:Pi_loop1}
\hat{\Pi}^{\mu \nu}(p)= 
U^{\mu \nu}(p) + \mathscr{R}^{\mu \nu}(0) + \mathcal{O}(a^\theta \absto{p}{\theta})
\end{equation}
where $U^{\mu \nu}(p)$ is the first term in~\eqref{eq:Pi_loop}.
The evaluation of the relativistic contribution $U^{\mu \nu}(p)$ 
is now straightforward due to the linearity of the dispersion relation, but one has
to take into account the presence of the momentum cutoff. It turns out that
\begin{equation}
\label{eq:Pi_matrix_free}
U^{\mu \nu}(p) 
= \frac{1}{4\pi} \sum_{c, \omega} Q^\textsc{a}_c Q^\textsc{v}_c 
\begin{pmatrix}
	\omega/\abs*{\mathfrak{v}_c}
&	 i  \\
	 i 
&	-\omega \abs*{\mathfrak{v}_c}	\\
\end{pmatrix}_{\! \mu \nu}
\frac{ip_0 + \omega \abs*{\mathfrak{v}_c} p_1}{-ip_0 + \omega \abs*{\mathfrak{v}_c} p_1} 
 + \mathcal{O}(a^\theta \absto{p}{\theta}).
\end{equation}
In order to see this, consider the identity
\begin{equation}
\label{eq:cutoff_WI}
\frac{\chi_b(q) \chi_b(p + q)}{D^+_b(q) D^+_b(p+q)} = \frac{1}{D^+_b(p)} \left(\frac{\chi_b(q)}{D^+_b(q)} - \frac{\chi_b(p + q)}{D^+_b(p+q)} \right) + \frac{\Delta_b(p, q)}{D^+_b(p)},
\end{equation}
where the correction term
\begin{equation}
\label{eq:Delta_definition}
\Delta_b(p, q) \equiv \frac{\chi_b(p+q)[1 - \chi_b(q)]}{D^+_b(p+q)} - \frac{\chi_b(q)[1 - \chi_b(p+q)]}{D^+_b(q)}
\end{equation}
is due to the fact that $\chi_b$ is not identically equal to $1$. 
We integrate both sides of~\eqref{eq:cutoff_WI} with respect to $q$, thus obtaining
\begin{equation}
\label{eq:integral_delta}
\int \frac{\dd[2]{q}}{(2\pi)^2} \, \frac{\chi_b(q) \chi_b(p + q)}{D^+_b(q) D^+_b(p+q)} = - \frac{1}{D^+_b(p)}\int \frac{\dd[2]{q}}{(2\pi)^2} \, \Delta_b(p, q)
\end{equation}
(the integral $\int \dd[2]{q} [\chi_b(q)/D_b(q) - \chi_b(p+q)/D_b(p+q)]$ vanishes). Note that the right hand side of~\eqref{eq:integral_delta} would be \emph{formally} vanishing if $\chi_b$ were replaced by $1$. By performing the change of variables $(q_0, q_1) \mapsto (q_0, q_1/\abs*{\mathfrak{v}_c})$ and subsequently writing $1/(-iq_0 + \omega q_1) = (iq_0 + \omega q_1)/\absto{q}{2}$,~\eqref{eq:integral_delta} becomes
\begin{equation}
\label{eq:integral_chi}
\int \frac{\dd[2]{q}}{(2\pi)^2} \frac{\chi_b(q) \chi_b(p + q)}{D^+_b(q) D^+_b(p+q)} = \frac{(i, \omega)_\mu}{4\pi^2 \abs*{\mathfrak{v}_c} D^+_b(p)} \int \dd[2]{q} \frac{q^\mu \chi(q)}{\absto{q}{2}} [\chi(q + p_c) - \chi(q-p_c)]
\end{equation}
with $p_c = (p_0, \abs*{\mathfrak{v}_c} p_1)$. We now note that $\chi(q \pm p_c) = \chi(q) \pm \absto{q}{-1}(q \cdot p_c) \chi'(q) + \int_0^1 \dd{t} p_c^\mu p_c^\nu \partial_\mu \partial_\nu \chi(q \pm p_c t)$, where $\chi'$ denotes the derivative of $\chi$ in the radial direction. The second derivative $\partial_\mu \partial_\nu \chi(q + p_c t)$ is of order $a^2$ and it is different from $0$ only provided that $\norm*{q} \sim 1/a$, so~\eqref{eq:integral_chi} can be rewritten as
\begin{equation}
\label{eq:integral_chi_quasi_final}
\int \frac{\dd[2]{q}}{(2\pi)^2} \frac{\chi_b(q) \chi_b(p + q)}{D^+_b(q) D^+_b(p+q)} = \frac{2p_{c, \nu} (i, \omega)_\mu}{4\pi^2 \abs*{\mathfrak{v}_c}\! D^+_b(p)} \int \dd[2]{q} \frac{q^\mu q^\nu}{\absto{q}{3}} \chi(q) \chi'(q) + \mathcal{O}(a \absto{p}{2}).
\end{equation}
Finally, we have
\begin{equation}
\int \dd[2]{q} \frac{q^\mu q^\nu}{\absto{q}{3}} \chi(q) \chi'(q) = \frac{\delta^{\mu \nu}}{2} \int \dd[2]{q} \frac{\chi(q) \chi'(q)}{\abs*{q}} = \frac{\delta^{\mu \nu}}{4} \int_0^{+\infty} 2\pi \dd{r} \frac{\dd}{\dd r} [\chi^2(r)] = -\frac{\pi \delta^{\mu \nu}}{2};
\end{equation}
by plugging this into~\eqref{eq:integral_chi_quasi_final}, we obtain
\begin{equation}
\label{eq:zeroth_order_cancellation}
\int \frac{\dd[2]{q}}{(2\pi)^2} \frac{\chi_{c, \omega}(q) \chi_{c, \omega}(p + q)}{D^+_{c, \omega}(q) D^+_{c, \omega}(p+q)} = \frac{1}{4\pi \abs*{\mathfrak{v}_c}} \frac{D^-_{c, \omega}(p)}{D^+_{c, \omega}(p)} + \mathcal{O}(a \absto{p}{2}).
\end{equation}
Formula~\eqref{eq:Pi_matrix_free} follows by replacing~\eqref{eq:zeroth_order_cancellation} inside~\eqref{eq:Pi_loop}. By a direct computation, one finds that
\begin{equation}
\label{eq:U_free}
ip_\alpha U^{\alpha \mu}(p) = 
-\frac{\epsilon^{\mu \alpha} p_\alpha}{2\pi} \sum_c Q^\textsc{a}_c Q^\textsc{v}_c.
\end{equation}
The function $ip_\alpha \mathscr{R}^{\alpha \mu}(p)$ explicitly
depends on all the lattice details. However, the exact validity of the vector current conservation
$$ip_\nu \hat{\Pi}^{\mu \nu}(p) = 0$$
implies that, as in the discussion after~\eqref{eq:contractions},
\begin{equation}
\label{eq:R_constant_free}
\begin{aligned}[c]
i\mathscr{R}^{\mu \nu}(0) = -\frac{\partial[ip_\alpha U^{\mu \alpha}(p)]}{\partial p_\nu} \eval_0 
& =
\frac{i}{4\pi} \sum_{c, \omega} Q^\textsc{a}_c Q^\textsc{v}_c \frac{\partial}{\partial p_\nu} \eval_0 
\begin{pmatrix}
\displaystyle \frac{(\omega p_0/\abs*{\mathfrak{v}_c}\! + ip_1)(-ip_0 - \omega \abs*{\mathfrak{v}_c} p_1)}{-ip_0 + \omega \abs*{\mathfrak{v}_c} p_1} \\[8pt]
\displaystyle \frac{(i p_0 - \omega \abs*{\mathfrak{v}_c} p_1)(-ip_0 - \omega \abs*{\mathfrak{v}_c} p_1)}{-ip_0 + \omega \abs*{\mathfrak{v}_c} p_1}
\end{pmatrix}_{\!\! \mu} \\
& = \frac{i}{4\pi} \sum_{c, \omega} Q^\textsc{a}_c Q^\textsc{v}_c
\begin{pmatrix}
\omega / \abs*{\mathfrak{v}_c}	&	-i \\
i						&	\omega \abs*{\mathfrak{v}_c}
\end{pmatrix}_{\! \mu \nu}
=
\frac{\epsilon^{\mu \nu}}{2\pi} \sum_c Q^\textsc{a}_c Q^\textsc{v}_c.
\end{aligned}
\end{equation}
Thanks to~\eqref{eq:R_constant_free} and~\eqref{eq:U_free}, we finally obtain
\begin{equation}
ip_\alpha [U^{\alpha \mu}(p) + \mathscr{R}^{\alpha \mu}(0)] = \frac{1}{2\pi} \sum_c Q^\textsc{a}_c Q^\textsc{v}_c (-\epsilon^{\mu \alpha} + \epsilon^{\alpha \mu}) p_\alpha = -\frac{ \epsilon^{\mu \alpha} p_\alpha}{\pi} \sum_c Q^\textsc{a}_c Q^\textsc{v}_c
\end{equation}
and consequently $ip_\alpha \hat{\Pi}^{\alpha \mu}(p) = -\pi^{-1}(\epsilon^{\mu \alpha} p_\alpha) \sum_c Q^\textsc{v}_c Q^\textsc{a}_c + \mathcal{O}(a^\theta \absto{p}{1 + \theta})$.

We emphasize that the contribution coming from the relativistic parts of the propagators with
momentum regularization both in space and time, encoded into the function 
$U^{\mu \nu}$, amounts to
half of the total chiral anomaly. The other half is provided by the constant $\mathscr{R}^{\mu\nu}(0)$, which accounts for irrelevant lattice effects. All the computation is based on the possibility of decomposing the correlation into the sum of a discontinuous part and a H\"{o}lder-continuous part; if $\hat{\Pi}^{\mu \nu}(p)$ were continuous, the same argument would produce a vanishing anomaly. Conversely, only the
H\"{o}lder-continuous part has a well-defined value at $p = 0$ which can be fixed by the continuity equation. The H\"{o}lder-continuous part contains terms that are irrelevant in the RG sense, but contribute to the anomaly as the relevant relativistic ones. 

In the main text, this strategy has been applied to the interacting case, where it is of course much more complicated. First of all, the correlations are expressed by a renormalized expansion in terms of running coupling constants, for which one needs to prove convergence.
One still wants to decompose the axial-vector correlation in a discontinuous part plus a continuous part. However, the discontinuous contribution corresponds to the axial-vector correlation of an \emph{interacting} relativistic QFT, not a free one, and this is true only if its parameters are suitably fine-tuned in a complicated way depending on the lattice details. Finally, one needs to compute exactly the correlation of this relativistic QFT showing that all the dependence on the lattice of the relativistic QFT disappears.

\subsection{II - Lowest order contributions to the renormalized expansion}
\label{suppl:renormalized_expansion1}
We write some lowest order contributions of the renormalized expansion to describe some
of the properties stated in the main text.

\paragraph{Interacting bubble.} We consider the first term in the first line of Figure~\ref{fig:first_few_diagrams}, whose value is given by
\begin{equation}
\label{eq:renormalized_bubble}
\begin{aligned}[c]
\begin{tikzpicture}[baseline={-0.6*height("$=$")}, scale=0.7]
\begin{feynman}
\draw[external photon]	(-1, 0)	--+(0.5, 0);
\draw[external photon]	(0.5, 0)	--+(0.5, 0);
\draw[with arrow]	(0.5, 0)		arc	(0:180:0.5);
\draw[with arrow]	(-0.5, 0)	arc	(180:360:0.5);
\node[scale=0.6]		(h)			at	($1.43*(90:0.5cm)$)		{$h$};
\node[scale=0.6]		(j)			at	($1.43*(-90:0.5cm)$)		{$j$};
\fill[black]	(-0.5, 0)	circle	(1.5pt);
\fill[black]	(0.5, 0)	circle	(1.5pt);
\end{feynman}
\end{tikzpicture}
& =
-\sum_b \frac{Z^\textsc{a}_{\mu, h, b}}{Z_{h, b}} \frac{Z^\textsc{v}_{\nu, h, b}}{Z_{h, b}} \frac{Z_{h, b}}{Z_{j, b}}
\begin{pmatrix}
\omega	&	i	\\
i		&	-\omega
\end{pmatrix}_{\! \mu \nu} 
\int \frac{\dd[2]{q}}{(2\pi)^2} \, \frac{f_{h, b}(q) f_{j, b}(q+p)}{D^+_{h, b}(q) D^+_{j, b}(q + p)} + \mathscr{R}^{\mu \nu}_{h, j}(p) \\
& \equiv \hat{B}^{\mu \nu}_{h, j}(p) + \mathscr{R}^{\mu \nu}_{h, j}(p),
\end{aligned}
\end{equation}
where $f_{h, b}(k) \equiv \chi_b(2^{-(h+1)} k) - \chi_b(2^{-h} k)$, $D^\pm_{h, b}(k) \equiv -ik_0 \pm \mathfrak{v}_{h, b} \, k_1$ and $\mathscr{R}^{\mu \nu}_{h, j}(p)$ is a remainder that accounts for lattice effects. In the above expression we have decomposed the lattice single scale propagator as
\begin{equation}
\label{eq:propagator_linearization}
\hat{g}^\ell_b(k) = \hat{\mathfrak{g}}^\ell_b(k) [1 + \hat{r}_{\ell, b}(k)], \qquad \hat{\mathfrak{g}}^\ell_b(k) \equiv \frac{f_{\ell, b}(k)}{Z_{\ell, b} D^+_{\ell, b}(k)}
\end{equation}
with $\abs*{\hat{r}_{\ell, b}(k)} \lesssim a^\theta\absto{k}{\theta}$ for some $\theta > 0$. The contributions coming from $\hat{r}_{\ell, b}(k)$ have been compressed into the remainder $\mathscr{R}^{\mu \nu}_{h, j}(p)$, which is H\"{o}lder-continuous. The above graph is different with respect to the non interacting bubble because of the presence of the renormalization constants $Z^\sharp_{\mu, h, b}, Z_{h, b}$; however, by
using Ward identities at each RG step as explained in~\cite{Mastropietro_Porta_2022}, one can show that
\begin{equation}
\label{eq:rcc_flow}
\begin{aligned}[c]
& \abs*{Z^\sharp_{0, h, b}/Z_{h, b} - Q^\sharp_c} \lesssim \lambda 2^{\theta h} \qquad \qquad 
& & \abs*{Z^\sharp_{1, h, b}/Z_{h, b} - \abs*{\mathfrak{v}_{-\infty, b}} Q^\sharp_c} \lesssim \lambda 2^{\theta h} \\
& Z_{h, b}/Z_{j, b} \simeq 2^{\eta_b(\lambda)(j - h)} \qquad \qquad 
& & \abs*{\mathfrak{v}_{h, b} - \mathfrak{v}_{-\infty, b}} \lesssim \lambda 2^{\theta h}
\end{aligned}
\end{equation}
with $\eta_b(\lambda) = \mathcal{O}(\lambda^2)$. By plugging~\eqref{eq:rcc_flow} inside~\eqref{eq:renormalized_bubble}, we obtain
\begin{equation}
\label{eq:interacting_bubble_dominant_part}
\begin{aligned}[c]
\sum_{h, j} \begin{tikzpicture}[baseline={-0.6*height("$=$")}, scale=0.7]
\begin{feynman}
\draw[external photon]	(-1, 0)	--+(0.5, 0);
\draw[external photon]	(0.5, 0)	--+(0.5, 0);
\draw[with arrow]	(0.5, 0)		arc	(0:180:0.5);
\draw[with arrow]	(-0.5, 0)	arc	(180:360:0.5);
\node[scale=0.6]		(h)			at	($1.43*(90:0.5cm)$)		{$h$};
\node[scale=0.6]		(j)			at	($1.43*(-90:0.5cm)$)		{$j$};
\fill[black]	(-0.5, 0)	circle	(1.5pt);
\fill[black]	(0.5, 0)	circle	(1.5pt);
\end{feynman}
\end{tikzpicture}
& = 
\begin{multlined}[t]
-\sum_b Q^\textsc{a}_c Q^\textsc{v}_c 
\begin{pmatrix}
	\omega
&	i \abs*{\mathfrak{v}_{-\infty, b}}  \\
	i \abs*{\mathfrak{v}_{-\infty, b}}
&	-\omega \absto{\mathfrak{v}_{-\infty, b}}{2}	\\
\end{pmatrix}_{\! \mu \nu}
\int \frac{\dd[2]{q}}{(2\pi)^2} \frac{\chi_{-\infty, b}(q) \chi_{-\infty, b}(q+p)}{D^+_{-\infty, b}(q) D^+_{-\infty, b}(q + p)} \, + \\
+ \sum_{b, h, j} \left(1 -  \frac{Z_{h, b}}{Z_{j, b}} \right) \int \frac{\dd[2]{q}}{(2\pi)^2} \frac{f_{h, b}(q) f_{j, b}(q + p)}{D^+_{h, b}(q) D^+_{j, b}(q + p)} + (\text{Continuous remainder}).
\end{multlined}
\end{aligned}
\end{equation}
Here, $\chi_{-\infty, b}$ and $D^+_{-\infty, b}$ contain the limit velocities $\mathfrak{v}_{-\infty, b}$. The first term at the right hand side of~\eqref{eq:interacting_bubble_dominant_part} is precisely equal to the non-continuous part of the non-interacting bubble displayed in~\eqref{eq:Pi_loop}, up to a renormalization of the Fermi velocities. Let us choose $h > j$ for definiteness; then, since $\abs*{1 - Z_{h, b}/Z_{j, b}} \lesssim \lambda^2 2^{\theta \abs*{h - j}}$ for some small $\theta > 0$, we have
\begin{equation}
\label{eq:loop_bound}
\abs{ \left(1 -  \frac{Z_{h, b}}{Z_{j, b}} \right) \! \int \frac{\dd[2]{q}}{(2\pi)^2} \frac{f_{h, b}(q) f_{j, b}(q + p)}{D^+_{h, b}(q) D^+_{j, b}(q + p)}} \lesssim \lambda^2 2^{\theta(h - j)} \cdot 2^{-h} \int \frac{\dd[2]{q}}{(2\pi)^2} \, \abs{\frac{f_{j, b}(q + p)}{D^+_{j, b}(q + p)}} \lesssim \lambda^2 2^{(1 - \theta)(j - h)}.
\end{equation}
Note that momentum conservation imposes the constraint $j \lesssim h_p \Rightarrow h \simeq h_p$, where $2^{h_p} = \abs*{p}$. Consequently,
\begin{equation}
\label{eq:sum_over_scales_bubble}
\begin{aligned}[c]
\sum_{h > j} \abs{\left(1 -  \frac{Z_{h, b}}{Z_{j, b}} \right) \int \frac{\dd[2]{q}}{(2\pi)^2} \frac{f_{h, b}(q) f_{j, b}(q+p)}{D^+_{h, b}(q) D^+_{j, b}(q + p)}}
& \lesssim \lambda^2 \sum_{j \lesssim h_p, h \simeq h_p} 2^{(1 - \theta)(j-h)} + \lambda^2 \sum_{j > h_p, h > j} 2^{(1 - \theta)(j-h)} \\
& \lesssim \lambda^2 [(\mathrm{const}) + h_p] \\
& \lesssim \lambda^2 \abs*{\log \abs*{p}\!}.
\end{aligned}
\end{equation}
A similar conclusion holds when $h < j$, so the first term in the second line of~\eqref{eq:interacting_bubble_dominant_part} is bounded by $\lambda^2 \abs*{\log \abs*{p}\!}\!$. This singular bound can however be improved thanks to cancellations: this is one of the main reasons for the introduction of the reference model in the main text.

\paragraph{A H\"{o}lder-continuous first-order diagram.}
Let us now analyze the first-order diagram occurring in the second line of Figure~\ref{fig:first_few_diagrams}. In this case there is a subdiagram over which the $\ren$ operation applies. The value of the diagram is
\begin{multline}
\label{eq:holder_continuous_diagram}
\begin{tikzpicture}[baseline={-0.6*height("$=$")}, scale=0.8]
\begin{feynman}
\draw[external photon]	(-1, 0)	--+(0.5, 0);
\draw[external photon]	(0.5, 0)	--+(0.5, 0);
\draw[photon]	(50:0.5cm)	to [out=240, in=120]	(-50:0.5cm);
\draw[with arrow]	(0.5, 0)			arc	(0:50:0.5);
\draw[with arrow]	(50:0.5cm)		arc	(50:180:0.5);
\draw[with arrow]	(-0.5, 0)		arc	(180:310:0.5);
\draw[with arrow]	(310:0.5cm)		arc	(310:360:0.5);
\draw[black, thick]	(0.1, -0.6)		rectangle	(0.8, 0.6);
\node[scale=0.8]		(a)	at	($1.4*(135:0.5cm)$)	{$h$};
\node[scale=0.8]		(b)	at	($1.45*(225:0.5cm)$)	{$j$};
\node[scale=0.8]		(c)	at	($1.45*(25:0.5cm)$)	{$k$};
\node[scale=0.8]		(d)	at	($1.4*(-25:0.5cm)$)	{$\ell$};
\end{feynman}
\end{tikzpicture}
= -\sum_b \int \frac{\dd[2]{q}}{(2\pi)^2} \hat{\mathfrak{g}}^h_b(q+p) \, u^\textsc{a}_{\mu, b, h} \, \hat{\mathfrak{g}}^j_b(q) \, \times \\
\times \ren \left [ \int \frac{\dd[2]{r}}{(2\pi)^2} \lambda_{\ell, bb} \, \hat{\mathfrak{g}}^\ell_{b}(q-r) \, u^\textsc{v}_{\nu, b, \ell} \, \hat{\mathfrak{g}}^k_{b}(q-r+p) \right] + \lambda \mathscr{R}_{\ell, k, h, j}^{\mu \nu}(p),
\end{multline}
where $\hat{\mathfrak{g}}^s_b(q) = f_{b, s}(q) [Z_{b, s} D^+_{b, s}(q)]^{-1}$, $u^\textsc{v}_{\mu, b, s} = (Z^\textsc{v}_{0, b, s}, i\omega Z^\textsc{v}_{1, b, s})_\mu$, $u^\textsc{a}_{\mu, b, s} = (\omega Z^\textsc{a}_{0, b, s}, i Z^\textsc{v}_{1, b, s})_\mu$ and $\mathscr{R}_{\ell, k, h, j}^{\mu \nu}(p)$ contains the lattice remainders (as before, the sum of this function over the scales $\ell, k, h, j$ is H\"{o}lder-continuous). The \emph{renormalization operator} $\ren$ has the effect of subtracting the local part of the marginal subdiagram evidenced with a black rectangle (this is necessary because the local parts of relevant and marginal subdiagrams are compressed into the running coupling constants).

Since the Grassmann variables anticommute, there is no such term as $\int \psi^+_b \psi^-_b \psi^+_b \psi^-_b$ inside the local part of $\mathscr{V}^\ell$, so $\lambda_{\ell, bb} \equiv 0$. This means that the whole first term appearing in~\eqref{eq:holder_continuous_diagram} vanishes, leading us to
\begin{equation}
\sum_{\substack{h, j \le 0 \colon\!\! \\[2pt] k, \ell \ge h, j}}
\left(
\begin{tikzpicture}[baseline={-0.6*height("$=$")}, scale=0.8]
\begin{feynman}
\draw[external photon]	(-1, 0)	--+(0.5, 0);
\draw[external photon]	(0.5, 0)	--+(0.5, 0);
\draw[photon]	(50:0.5cm)	to [out=240, in=120]	(-50:0.5cm);
\draw[with arrow]	(0.5, 0)			arc	(0:50:0.5);
\draw[with arrow]	(50:0.5cm)		arc	(50:180:0.5);
\draw[with arrow]	(-0.5, 0)		arc	(180:310:0.5);
\draw[with arrow]	(310:0.5cm)		arc	(310:360:0.5);
\draw[black, thick]	(0.1, -0.6)		rectangle	(0.8, 0.6);
\node[scale=0.8]		(a)	at	($1.4*(135:0.5cm)$)	{$h$};
\node[scale=0.8]		(b)	at	($1.45*(225:0.5cm)$)	{$j$};
\node[scale=0.8]		(c)	at	($1.45*(25:0.5cm)$)	{$k$};
\node[scale=0.8]		(d)	at	($1.4*(-25:0.5cm)$)	{$\ell$};
\end{feynman}
\end{tikzpicture}
\right)
=
\sum_{\substack{h, j \le 0 \colon\!\! \\[2pt] k, \ell \ge h, j}} \lambda \mathscr{R}_{\ell, k, h, j}^{\mu \nu}(p) \equiv \lambda \mathscr{R}^{\mu \nu}(0) + \mathcal{O}(\lambda \absto{p}{\theta}).
\end{equation}
Hence, this diagram is H\"{o}lder-continuous. We stress that even if its relativistic part vanishes, this diagram \emph{does} contribute to the dominant part of the lattice chiral anomaly with the linear term $ip_\alpha[\lambda \mathscr{R}^{\alpha \mu}(0)]$.

\paragraph{A discontinuous second-order diagram.}
We finally consider a second-order diagram whose relativistic part is discontinuous at $p = 0$, namely
\begin{equation}
\label{eq:three_loop_diagram}
\begin{tikzpicture}[baseline={-0.6*height("$=$")}, scale=0.8]
\begin{feynman}
\draw[red, thick]				(45:1cm)	arc	(45:135:1);
\draw							(225:1cm)	arc	(225:315:1);
\draw[with arrow, red, thick]	(-45:1cm)	arc	(-45:0:1);
\draw[with arrow, red, thick]	(0:1cm)		arc	(0:45:1);
\draw[with arrow, red, thick]	(135:1cm)	arc	(135:180:1);			
\draw[with arrow]	(180:1cm)	arc	(180:225:1);
\draw[with arrow]	(-45:0.3cm)	arc	(-45:45:0.3);
\draw[with arrow]	(45:0.3cm)	arc	(45:315:0.3);
\draw[photon]		(45:1cm)	--	(45:0.3cm);
\draw[photon]		(-45:1cm)	--	(-45:0.3cm);
\draw[external photon]	(-1, 0)	--	(-1.5, 0);
\draw[external photon]	(1, 0)	--	(1.5, 0);
\draw[black, thick]		(-0.4, -1.1) rectangle (1.3, 1.1);
\node[scale=0.8]	at	(135:1.2cm)		{$h$};
\node[scale=0.8]	at	(225:1.2cm)		{$j$};
\node[scale=0.8]	at	(180:0.1cm)		{$r$};
\node[scale=0.8]	at	(0:0.5cm)		{$s$};
\node[scale=0.8]	at	(25:1.2cm)		{$k$};
\node[scale=0.8]	at	(-25:1.2cm)		{$\ell$};
\end{feynman}
\end{tikzpicture}
\equiv \hat{G}^{\mu \nu}_{j, \dots, \ell}(p) + \mathscr{R}^{\mu \nu}_{j, \dots, \ell}(p)
\end{equation}
with $\ell > k > r > s > h > j$. Here, $\hat{G}^{\mu \nu}_{j, \dots, \ell}(p)$ denotes the relativistic part of the diagram, that is
\begin{multline}
\label{eq:relativistic_two_loops}
\hat{G}^{\mu \nu}_{j, \dots, \ell}(p) = \sum_b \int \frac{\dd[2]{q_1}}{(2\pi)^2} \, \hat{\mathfrak{g}}^h_b(q_1 + p) \, u^\textsc{a}_{h, \mu, b} \, \hat{\mathfrak{g}}^j_b(q_1) \times \\
\times \ren \left[ \sum_{b'} \int \frac{\dd[2]{q_2}}{(2\pi)^2} \, \lambda_{\ell, bb'} \hat{\mathfrak{g}}^\ell_b(q_1 - q_2) \, u^\textsc{v}_{\ell, \nu, b} \, \hat{\mathfrak{g}}^k_b(q_1 - q_2 + p) \lambda_{k, bb'} \!\! \int \frac{\dd[2]{q_3}}{(2\pi)^2} \, \hat{\mathfrak{g}}^r_{b'}(q_3 - q_2) \hat{\mathfrak{g}}^s_{b'}(q_3) \right],
\end{multline} 
whereas $\mathscr{R}^{\mu \nu}_{j, \dots, \ell}(p)$ is a continuous lattice remainder.

Let us derive a dimensional bound for $\hat{G}^{\mu \nu}_{j, \dots, \ell}(p)$. This can be conveniently done by estimating the position-space integral $\int \dd{x} \abs*{G^{\mu \nu}_{j, \dots, \ell}(x)}\!$. Consider the tree subgraph $T$ evidenced in red inside~\eqref{eq:three_loop_diagram}; since $T$ is a tree graph, the integral over all the spacetime points occurring inside the diagram can be recast as an integral over the differences of spacetime points lying at the ends of the edges of $T$. Consequently, all the propagators lying along $T$ contribute to the bound as $\int \dd{x} \abs*{g^{h_a}(x)}\! \lesssim 2^{-h_a}$ and all the remaining propagators are instead bounded as $\abs*{g^{h_a}(x)}\! \lesssim 2^{h_a}$. Since the subdiagram $\hat{\kappa}(\underline{t})$ evidenced with a black rectangle is deprived of its local part, it contributes to $\hat{G}^{\mu \nu}_{j, \dots, \ell}(p)$ through the combination $\hat{\kappa}(\underline{t}) - \hat{\kappa}(\underline{0}) = \underline{t} \cdot \int_0^1 \dd{u} \underline{\partial} \hat{\kappa}(u\underline{t})$. The momenta $\underline{t}$ are supported below the scale $h$, while the derivatives $\underline{\partial}$ fall on the internal propagators of $\hat{\kappa}$, which are supported above the scale $s$; therefore, the renormalization of $\hat{\kappa}$ produces an extra $2^{h-s}$ gain factor. In summary, we obtain
\begin{multline}
\label{eq:dimensional_bound_three_loops}
\abs*{\hat{G}^{\mu \nu}_{j, \dots, \ell}(p)} \le \int \dd{x} \abs*{G^{\mu \nu}_{j, \dots, \ell}(x)} \lesssim \lambda^2 \cdot \underbrace{2^{-h-k-j}}_{\prod_{a \in \mathrm{E}(T)} \norm*{g^{h_a}\!}_1} \,\,\,\, \cdot \underbrace{2^{r + s + j}}_{\prod_{a \notin \mathrm{E}(T)} \norm*{g^{h_a}\!}_\infty} \,\, \cdot \!\! \underbrace{2^{h - s}}_{\textup{Renormalization}} \, = \\
= \lambda^2 2^{-(\ell - k)} 2^{-2(k - r)} 2^{-(r - s)} 2^{-(s - h)} 2^{-(h - j)}.
\end{multline}
The sum over the scales can be done as in~\eqref{eq:sum_over_scales_bubble}. The result is that $\sum_{j < \dots < \ell} \abs*{\hat{G}^{\mu \nu}_{j, \dots, \ell}(p)} \lesssim \lambda^2 \abs*{\log \abs*{p}\!}\!$, so dimensional arguments suggest the existence of a logarithmic singularity.

\subsection{III - Improvement of the logarithmic behavior}

The problem of the apparent logarithmic divergence of $\hat{\Pi}^{\mu \nu}(p)$ is solved in the main text by introducing the reference model. According to~\eqref{eq:Pi_reference_model_comparison}, the correlation $\hat{\Pi}^{\mu \nu}(p)$ can be written as the sum of its reference model counterpart, which has an \emph{explicit} form that can be deduced from the Ward Identities, plus a remainder 
which is finite and  H\"{o}lder-continuous.
Here we shall see why this is true in the interacting bubble. (The exact same argument can be applied to the second-order diagram~\eqref{eq:three_loop_diagram} as well).

Let $\hat{B}^{\mu \nu}_{\rff, h, j}(p)$ be the reference model version of the interacting bubble. Then, we can write
\begin{equation}
\label{eq:bubble_reference_splitting}
\begin{aligned}[c]
\begin{tikzpicture}[baseline={-0.6*height("$=$")}, scale=0.7]
\begin{feynman}
\draw[external photon]	(-1, 0)	--+(0.5, 0);
\draw[external photon]	(0.5, 0)	--+(0.5, 0);
\draw[with arrow]	(0.5, 0)		arc	(0:180:0.5);
\draw[with arrow]	(-0.5, 0)	arc	(180:360:0.5);
\node[scale=0.6]		(h)			at	($1.43*(90:0.5cm)$)		{$h$};
\node[scale=0.6]		(j)			at	($1.43*(-90:0.5cm)$)		{$j$};
\fill[black]	(-0.5, 0)	circle	(1.5pt);
\fill[black]	(0.5, 0)	circle	(1.5pt);
\end{feynman}
\end{tikzpicture}
= \hat{B}^{\mu \nu}_{\rff, h, j}(p) + [\hat{B}^{\mu \nu}_{h, j}(p) - \hat{B}^{\mu \nu}_{\rff, h, j}(p)] +  \mathscr{R}^{\mu \nu}_{h, j}(p).
\end{aligned}
\end{equation}
The key fact about this decomposition is that the function $F^{\mu \nu}(p) = \sum_{h, j} [\hat{B}^{\mu \nu}_{h, j}(p) - \hat{B}^{\mu \nu}_{\rff, h, j}(p)]$ is H\"{o}lder-continuous. To see this, let us start by proving that $F^{\mu \nu}(p)$ is finite at $p = 0$. The difference $\hat{B}^{\mu \nu}_{h, j}(p) - \hat{B}^{\mu \nu}_{\rff, h, j}(p)$ has the same structure as $\hat{B}^{\mu \nu}_{h, j}(p)$, except for the fact that it contains at least one $Z^\sharp_{s, b} - \mathcal{Z}^\sharp_{s, b}$ factor in place of $Z^\sharp_{s, b}$ and/or a $f_{\rff, s, b}(q)/[\mathcal{Z}_{s, b} D_{\rff, s, b}^+(q)] - f_{s, b}(q)/[Z_{s, b} D_{s, b}^+(q)]$ factor in place of $f_{s, b}(q)/[Z_{s, b} D_{s, b}^+(q)]$. Since the reference model and the lattice model share the same infrared fixed point, these differences produce an extra $\smash{2^{\theta \max \lbrace h, j \rbrace}}$ gain with respect to the usual dimensional bounds. If $h > j$, we have
\begin{equation}
\label{eq:sum_over_scales_bubble_bound}
\abs{\sum_{h > j} [\hat{B}^{\mu \nu}_{h, j}(p) - \hat{B}^{\mu \nu}_{\rff, h, j}(p)]} \lesssim \sum_{j \lesssim h_p, h \simeq h_p} 2^{j-h + \theta h} + \sum_{j > h_p, h > j} 2^{j-h + \theta h} \lesssim \absto{p}{\theta} + (\mathrm{const}) \lesssim (\mathrm{const}).
\end{equation}
The same is true for $h < j$, so $F^{\mu \nu}(p)$ is finite at $p = 0$. 

The H\"{o}lder-continuity of $F^{\mu \nu}(p)$ follows from an analogous argument. The difference $F^{\mu \nu}(p) - F^{\mu \nu}(0)$ is given by a sum of terms with the same form as those contributing to $F^{\mu \nu}(p)$, except that they contain at least one $f_{j, b}(k)/D^+_{j, b}(k) - f_{j, b}(k+p)/D^+_{j, b}(k+p)$ factor or one $f_{\rff, j, b}(k)/D^+_{\rff, j, b}(k) - f_{\rff, j, b}(k+p)/D^+_{\rff, j, b}(k+p)$ factor. If $j \gtrsim h_p, h \ge j$, it is convenient to represent these factors as $p \cdot \partial[f_{j, b}/D^+_{j, b}]$ or $p \cdot \partial[f_{\rff, j, b}/D^+_{\rff, j, b}]$; therefore, the bound for $F^{\mu \nu}(p) - F^{\mu \nu}(0)$ is similar to~\eqref{eq:sum_over_scales_bubble_bound}, with the exception that it contains an extra $\abs*{p} \cdot 2^{-j}$ factor in the $j \gtrsim h_p, h \ge j$ regime. The result is
\begin{align*}
\abs*{F^{\mu \nu}(p) - F^{\mu \nu}(0)}
& \lesssim \sum_{j \lesssim h_p, h \simeq h_p} \lambda 2^{j - h + \theta h} + \sum_{j \gtrsim h_p, h \ge j} \abs*{p} \cdot \lambda 2^{j - h + \theta h} \cdot 2^{-j} \\
& = \sum_{j \lesssim h_p, h \simeq h_p} \lambda 2^{j - h + \theta h} + \absto{p}{\theta} \sum_{j \gtrsim h_p, h \ge j} 2^{(1 - \theta)(h_p - j)} \cdot \lambda 2^{(1 - \theta)(j - h)} \\
& \lesssim \absto{p}{\theta}
\end{align*}
and this implies that $F^{\mu \nu}(p)$ is H\"{o}lder-continuous at $p = 0$. In summary, we showed that~\eqref{eq:bubble_reference_splitting} agrees with~\eqref{eq:Pi_reference_model_comparison} at a one-loop level.

\subsection{IV - Ward Identities for the reference model}
\label{suppl:Ward_Identities}

Here we briefly explain how to derive~\eqref{eq:anomalous_WI_Gamma} and~\eqref{eq:anomalous_WI_rho} when $\lambda^\rff_{bb'} \ne 0$. A complete, rigorous treatment of this problem can be found in~\cite{Mastropietro_Porta_2022}. Formulae~\eqref{eq:anomalous_WI_Gamma},~\eqref{eq:anomalous_WI_rho} are obtained by performing the change of variables $\psi^\pm_{b'}(x) \mapsto e^{\pm i \delta_{bb'} \alpha(x)} \psi^\pm_{b'}(x)$ inside the functional integral $\int \dgauss{g}{\psi} \, e^{- V^\rff[\smash{\sqrt{\mathcal{Z}}} \psi] + (\mathrm{sources})}$. In Fourier space, the phase transformation reads $\hat{\psi}^{\pm}_{b', k} \mapsto \hat{\psi}^{\pm}_{b', k} \pm i \delta_{bb'} L^{-2}\sum_q \hat{\alpha}_{\pm(k - q)} \hat{\psi}^{\pm}_{b, q} + \mathcal{O}(\alpha^2)$, so the free action $S_0[\psi] = L^{-2} \sum_{b, k} \hat{\psi}^+_{b, k} \hat{\psi}^-_{b, k} \mathcal{Z}_b D^+_b(k) \chi^{-1}_b(k)$ transforms as
\begin{align*}
e^{-S_0[\psi]} \mapsto
& \, e^{-S_0[\psi]} \left( 1 - \frac{i\mathcal{Z}_b}{L^4} \sum_{k, q} [\hat{\psi}^+_{b, q} \hat{\alpha}_{k - q} \hat{\psi}^-_{b, k} - \hat{\psi}^+_{b, k} \hat{\alpha}_{q - k} \hat{\psi}^-_{b, q}] D_b^+(k) \chi^{-1}_b(k) + \mathcal{O}(\alpha^2) \right) \\
= & \, e^{-S_0[\psi]} \biggl( 1 - \frac{i\mathcal{Z}_b}{L^4} \sum_{k, q} \hat{\alpha}_k \hat{\psi}^+_{b, q} \hat{\psi}^-_{b, k+q} [D_b^+(k+q) \chi^{-1}_b(k+q) - D_b^+(q) \chi^{-1}_b(q)] + \mathcal{O}(\alpha^2) \biggr).
\end{align*}
The term that multiplies $\hat{\alpha}_k$ can be written as $-i D_b^+(k) \mathcal{Z}_b \hat{\psi}^+_{b, q} \hat{\psi}^-_{b, k+q} + i \mathcal{Z}_b \hat{\psi}^+_{b, q} \hat{\psi}^-_{b, k+q} C_b(k, q)$, with
\begin{equation}
C_b(k, q) = D_b^+(k+q)(1 - \chi^{-1}_b(k+q)) - D_b^+(q)(1 - \chi^{-1}_b(q));
\end{equation}
therefore, since $\hat{\rho}_{b, k} = L^{-2} \sum_q \hat{\psi}^+_{b, q - k} \hat{\psi}^-_{b, q}$, the phase transformation acts on the Gaussian Grassmann measure as
\begin{equation}
\label{eq:free_measure_transformation}
\dgauss{g^\rff}{\psi} \mapsto \dgauss{g^\rff}{\psi} \left[ 1 - \frac{i}{L^2} \sum_k \hat{\alpha}_k D^+_b(k) \mathcal{Z}_b \hat{\rho}_{b, k} + \frac{i\mathcal{Z}_b}{L^4} \sum_{k, q} \hat{\alpha}_k \hat{\psi}^+_{b, q} \hat{\psi}^-_{b, k+q} C_b(k, q) + \mathcal{O}(\alpha^2) \right]
\end{equation}
whereas the source terms $\mathcal{B}[\psi, \phi, K, J^\sharp] \equiv \sum_b \int (\psi^+_b \phi^-_b + \phi^+_b \psi^-_b + K_b \rho_b) + \sum_\sharp \int J^\sharp \cdot j^\sharp$ transform as
\begin{equation}
\label{eq:sources_transformation}
e^{\mathcal{B}[\psi, \phi, K, J^\sharp]} \mapsto e^{\mathcal{B}[\psi, \phi, K, J^\sharp]} \left[1 + \frac{i}{L^4} \sum_{k, q} \hat{\alpha}_k (\hat{\psi}^+_{b, q} \hat{\phi}^-_{b, k + q} - \hat{\phi}^+_{b, q} \hat{\psi}^-_{b, k + q}) + \mathcal{O}(\alpha^2) \right];
\end{equation}
finally, the interaction term $V^\rff[\smash{\sqrt{\mathcal{Z}}}\psi]$ is clearly left unchanged. The functional integral is globally unaffected by this change of variables for every choice of $\alpha$, so its partial derivative with respect to $\hat{\alpha}_p$ must vanish for every $p$. By looking at~\eqref{eq:free_measure_transformation} and~\eqref{eq:sources_transformation}, we deduce that this condition is satisfied if
\begin{equation}
\label{eq:formal_WI_generator}
-D^+_b(p) \mathcal{Z}_b \expval*{\hat{\rho}_{b, p}}_\textsc{s} + \frac{\mathcal{Z}_b}{L^2} \sum_q C_b(p, q) \expval*{\hat{\psi}^+_{b, q} \hat{\psi}^-_{b, p+q}}_\textsc{s} + \frac{1}{L^2} \sum_q \expval*{\hat{\psi}^+_{b, q} \hat{\phi}^-_{b, p + q} - \hat{\phi}^+_{b, q} \hat{\psi}^-_{b, p + q}}_\textsc{s} = 0,
\end{equation}
where $\expval*{ \, \cdot \, }_{\textsc{s}}$ denotes a vacuum expectation value in presence of nonvanishing $\phi, J^\sharp, K$ source fields. If the second term at the left hand side of~\eqref{eq:formal_WI_generator} were absent, we could take suitable derivatives with respect to $\phi^+, \phi^-, K$ and subsequently obtain the formal Ward Identities
\begin{equation}
\label{eq:formal_WIs}
D^+_{b}(p) \mathcal{Z}_b \hat\Gamma_{b b'}^\rff(k, p) = \delta_{bb'} [\hat{S}_b(k) - \hat{S}_b(k+p)], \qquad
D^+_{b}(p) \mathcal{Z}_b \hat{A}_{b b'}^\rff(p) = 0,
\end{equation}
which simply express the conservation of the currents $j^\mu_b = \mathcal{Z}_b \rho_b(1, i\mathfrak{u}_b)_\mu$ in the usual sense. However, this is not the case: the actual Ward Identities of the reference model are \emph{anomalous}, because $\chi_b$ is not identically $1$ and consequently $C_b(p, q) \ne 0$. The effects of $C_b(p, q)$ are graphically represented by the last two graphs occurring in each line of Figure~\ref{fig:anomalous_WI}.

The precise form of the corrections coming from $C_b(p, q)$ has been computed in~\cite{Mastropietro_Porta_2022}. Here we can at least qualitatively understand the structure of these corrections in a non-rigorous, ordinary perturbative context. For the sake of simplicity, let us consider the vertex Ward Identity, which follows from differentiating~\eqref{eq:formal_WI_generator} with respect to $\phi^+, \phi^-$. We divide all the possible Feynman diagrams that can contribute to the correction term as
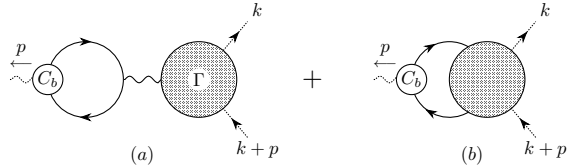
\begin{center}
\begin{tikzpicture}[baseline]
\begin{feynman}
\draw[external photon]	(-0.5, 0) -- (0, 0);
\draw[with arrow]		(0, 0)		arc	(180:0:0.5);
\draw[with arrow]		(1, 0)		arc	(0:-180:0.5);
\draw[photon]			(1, 0)	--	(1.5, 0);
\draw[external fermion]	($(2, 0) + (-50:1cm)$) 	-- 	($(2, 0) + (-50:0.5cm)$);
\draw[external fermion]	($(2, 0) + (50:0.5cm)$) 	-- 	($(2, 0) + (50:1cm)$);
\fill[draw = black, correlator background]	(2, 0)	circle	(0.5cm);
\fill[white, draw = black]		(0, 0)	circle	(0.2cm);
\node[circle, scale=0.7, fill=white, text=black,%
		inner sep=0, minimum size=1.2em]		(C)		at	(0, 0)	{$C_b$};
\node[circle, scale=0.7, fill=white, text=black,%
		inner sep=0, minimum size=1.2em]		(G)		at	(2, 0)	{$\Gamma$};
\node				(plus)	at	(3.5, 0)			{$+$};
\node[scale=0.7]		(a)		at	(1.25, -1)	{$(a)$};
\node[scale=0.7]		(kp)		at	($(2, 0) + (-50:1.2cm)$)	{$k+p$};
\node[scale=0.7]		(k)		at	($(2, 0) + (50:1.2cm)$)	{$k$};
\node[scale=0.7]		(p)		at	(-0.35, 0.3)	{$\underset{\longleftarrow}{p}$};
\end{feynman}
\end{tikzpicture}
\quad
\begin{tikzpicture}[baseline]
\begin{feynman}
\draw[external photon]	(-0.5, 0) -- (0, 0);
\draw[with arrow]		(0, 0)		arc	(180:60:0.5);
\draw[with arrow]		($(0.5, 0) + (-60:0.5cm)$)	arc	(-60:-180:0.5);
\draw[external fermion]	($(1, 0) + (-50:1cm)$) 	-- 	($(1, 0) + (-50:0.5cm)$);
\draw[external fermion]	($(1, 0) + (50:0.5cm)$) 	-- 	($(1, 0) + (50:1cm)$);
\fill[draw = black, correlator background]	(1, 0)		circle	(0.5cm);
\fill[white, draw = black]		(0, 0)	circle	(0.2cm);
\node[circle, scale=0.7, fill=white, text=black,%
		inner sep=0, minimum size=1.2em]		(C)		at	(0, 0)	{$C_b$};
\node[scale=0.7]		(b)		at	(0.8, -1)	{$(b)$};
\node[scale=0.7]		(kp)		at	($(1, 0) + (-50:1.2cm)$)	{$k+p$};
\node[scale=0.7]		(k)		at	($(1, 0) + (50:1.2cm)$)	{$k$};
\node[scale=0.7]		(p)		at	(-0.35, 0.3)	{$\underset{\longleftarrow}{p}$};
\end{feynman}
\end{tikzpicture}
\end{center}
where 
\begin{itemize}
\item Diagrams of type $(a)$ are obtained by contracting the two fermionic lines emerging from the kernel $C_b(p, q)$ into a non-interacting bubble, which is then connected with the two fermionic external legs in all the possible ways.
\item Type $(b)$ encompasses all the diagrams that are not of type $(a)$. In a diagram of type $(b)$, the kernel $C_b(p, q)$ is \emph{not} contracted with a non-interacting bubble.
\end{itemize}
The sum of all the diagrams of type $(a)$ is equal to $-F_b(p) \hat{v}_p \sum_{b'} \lambda^\rff_{bb'} \mathcal{Z}_{b'} \hat{\Gamma}_{b'b}(k, p)$, where 
\begin{equation}
F_b(p) = \int \frac{\dd[2]{q}}{(2\pi)^2} \hat{g}^\rff_b(q) \hat{g}^\rff_b(p + q) C_b(p, q) = \int \frac{\dd[2]{q}}{(2\pi)^2} \Delta_b(p, q).
\end{equation}
Thanks to~\eqref{eq:integral_delta} and~\eqref{eq:zeroth_order_cancellation}, we have $F_b(p) = -D^-_b(p)(4\pi \abs*{\mathfrak{u}_b}\!)^{-1}$ up to $\mathcal{O}(2^{-N}\absto{p}{2})$ corrections. Therefore, the sum of all the diagrams of type $(a)$ contributes to the vertex Ward Identities with the anomalous term $D^-_b(p)(4\pi \abs*{\mathfrak{u}_b}\!)^{-1} \hat{v}_p \sum_{b'} \lambda^\rff_{bb'} \mathcal{Z}_{b'} \hat{\Gamma}_{b'b}(k, p)$, which coincides with the last graph displayed in the second line of Figure~\ref{fig:anomalous_WI} and is encoded into the non-diagonal part of the matrix $T^{-1}_{bb'}(p)$ inside~\eqref{eq:anomalous_WI_Gamma}.

All the diagrams of type $(b)$ \emph{vanish} when the ultraviolet limit is taken, so the anomaly is entirely determined by the diagrams of type $(a)$ analyzed above. The proof of this property is based on the key fact that the kernel $\hat{g}^\rff_b(q) \hat{g}^\rff_b(p + q)C_b(p, q)$ behaves as $D^+_b(p) \hat{g}^j(q) \hat{g}^k(p+q)$ with either $j \sim N$ or $k \sim N$. This makes it possible to extract an overall $\mathcal{O}(2^{-\theta N})$ suppression factor that causes any diagram of type $(b)$ to vanish as $N \to +\infty$. To see a concrete example of this, let us consider the type-$(b)$ diagram
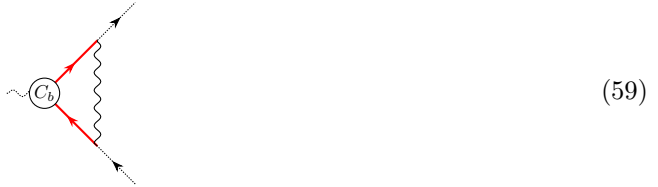
\begin{equation}
\label{eq:triangle_graph}
\begin{tikzpicture}[baseline={-0.6*height("$=$")}]
\begin{feynman}
\draw[external photon]	(-0.5, 0) -- (0, 0);
\draw[with arrow, red, thick]	(0.7, -0.7)	--	(0, 0);
\draw[with arrow, red, thick]	(0, 0)	--	(0.7, 0.7);
\fill[white, draw = black]		(0, 0)	circle	(0.2cm);
\node[circle, scale=0.7, fill=white, text=black,%
		inner sep=0, minimum size=1.2em]		(C)		at	(0, 0)	{$C_b$};
\draw[photon]			(0.7, 0.7)	--	(0.7, -0.7);
\draw[external fermion]	(0.7, 0.7)	--	(1.2, 1.2);
\draw[external fermion]	(1.2, -1.2)		--	(0.7, -0.7);
\end{feynman}
\end{tikzpicture}
\end{equation}
and suppose that the momentum flowing through one of the two fermion lines is of order $2^N$. Due to momentum conservation, the momentum flowing through the other fermion line must be of order $2^N$ as well if $\abs*{p}\!$ is sufficiently small with respect to the ultraviolet cutoff. A dimensional bound for this diagram can be found with the same technique discussed before~\eqref{eq:dimensional_bound_three_loops}: this time, since $\int \dd{x} \abs*{g^N}\! \lesssim 2^{-N}$ and $\abs*{v(x)}\! \le (\mathrm{const})$, it is convenient to choose the tree subdiagram $T$ evidenced in red inside~\eqref{eq:triangle_graph}. As a result, one sees that the diagram is bounded by $\abs*{D^+_b(p)} \cdot 2^{-N - N} \cdot (\mathrm{const}) \overset{N \to +\infty}{\longrightarrow} 0$. 

In order to prove that this diagram vanishes as $N \to +\infty$, we crucially exploited the fact that the two fermion lines attached to $C_b(p, q)$ do \emph{not} close into a single non-interacting bubble. This allows to include both these lines inside the tree $T$, thus yielding a $2^{-2N}$ dimensional gain. If we repeated the same reasoning with a diagram of type $(a)$, we would be forced to exclude one of the two fermion lines attached to $C_b(p, q)$ from the tree $T$: this would produce a much worse $\mathcal{O}(2^{-N} \cdot 2^N) = \mathcal{O}(1)$ bound.

\subsection{V - Check of Equations \texorpdfstring{\eqref{eq:constraints},~\eqref{eq:contractions} and~\eqref{eq:M_matrix}}{e}}
\label{subsuppl:computations}

We provide here some details on the formulae appearing in the paper. 

\paragraph{Proof of~\eqref{eq:constraints}.} In order to prove~\eqref{eq:constraints}, we write the Ward Identity~\eqref{eq:anomalous_WI_Gamma} in matrix form as $T^{-1} \mathcal{Z} \hat{\Gamma}^\rff = \Delta \hat{S}^\rff (D^+)^{-1}$, where $\mathcal{Z}$, $\Delta \hat{S}^\rff$, $D^+$ are thought of as diagonal matrices having $\mathcal{Z}_b$, $\Delta \hat{S}^\rff_b$, $D^+_b$ as nonzero entries. A simple inversion yields $\hat{\Gamma}^\rff = \mathcal{Z}^{-1} T \Delta \hat{S}^\rff (D^+)^{-1}$; therefore, since $\hat{\Gamma}^{\rff, \sharp}_{\mu, b'} = \sum_b \sigma^\sharp_{\mu, b} \hat{\Gamma}^\rff_{bb'}$, we have $
\hat{\Gamma}^{\rff, \sharp}_{\mu, b'}(p, p) = \sum_b \sigma^\sharp_{\mu, b} T_{bb'}(p) \Delta \hat{S}^\rff_{b'}(p, p)[\mathcal{Z}_b D^+_{b'}(p)]^{-1}$. The reference vertex $
\hat{\Gamma}^{\rff, \sharp}_{\mu, c'}(p, p)$ is equal to 
$\hat{\Gamma}^\sharp_{\mu, c'}(\bar{p}_{b'}, p)$ up to subleading terms, so
\begin{equation}
\label{eq:vertices_matching}
i p_\mu \hat{\Gamma}^\sharp_{\mu, c'}(\bar{p}_{b'}, p) = \sum_b  i p_\mu \sigma^\sharp_{\mu, b} \frac{T_{bb'}(p) \Delta \hat{S}^\rff_{b'}(p, p)}{\mathcal{Z}_b D^+_{b'}(p)} + \mathcal{O}(\absto{p}{\theta - 1}).
\end{equation}
If $\sharp = \textsc{v}$, the left hand side of~\eqref{eq:vertices_matching} satisfies the lattice Ward Identity~\eqref{eq:lattice_WI_vertex}, so it must be 
\begin{equation}
\sum_b  i p_\mu \sigma^\sharp_{\mu, b} \frac{T_{bb'}(p) \Delta \hat{S}^\rff_{b'}(p, p)}{\mathcal{Z}_b D^+_{b'}(p)} = -Q_{c'}^\textsc{v}[\hat{S}_{c'}(\overline{p}_{b'}) - \hat{S}_{c'}(\overline{2p}_{b'})] + \mathcal{O}(\absto{p}{\theta - 1}).
\end{equation}
Since $\hat{S}_{c'}(\overline{p}_{b'}) - \hat{S}_{c'}(\overline{2p}_{b'})$ coincides with $\Delta \hat{S}^\rff_{b'}(p, p) = \mathcal{O}(\absto{p}{-[1 + \mathcal{O}(\lambda^2)]})$ up to subleading terms, this equality reduces to formula~\eqref{eq:constraints} with $\sharp = \textsc{v}$. To derive~\eqref{eq:constraints} in the $\sharp = \textsc{a}$ case, we note that the axial charge normalization condition~\eqref{eq:normalization_condition} implies that $\hat{\Gamma}^\textsc{a}_{\mu, c'}(\bar{p}_{b'}, p)$ is equal to $(\omega' Q^\textsc{a}_{c'}/Q^\textsc{v}_{c'}) \hat{\Gamma}^\textsc{v}_{\mu, c'}(\bar{p}_{b'}, p)$ up to subleading terms, so~\eqref{eq:vertices_matching} with $\sharp = \textsc{a}$ becomes
\begin{equation}
\label{eq:axial_vertex_matching}
i p_\mu \hat{\Gamma}^\textsc{v}_{b'}(\bar{p}_{b'}, p) \cdot \frac{\omega' Q^\textsc{a}_{c'}}{Q^\textsc{v}_{c'}} = \sum_b  i p_\mu \sigma^\textsc{a}_{\mu, b} \frac{T_{bb'}(p) \Delta \hat{S}^\rff_{b'}(p, p)}{\mathcal{Z}_b D^+_{b'}(p)} + \mathcal{O}(\absto{p}{\theta - 1}).
\end{equation}
The lattice Ward Identity~\eqref{eq:lattice_WI_vertex} applied to the left hand side of~\eqref{eq:axial_vertex_matching} yields $\sum_b  i p_\mu \sigma^\textsc{a}_{\mu, b} T_{bb'}(p) \allowbreak  \Delta \hat{S}^\rff_{b'}(p, p)[\mathcal{Z}_b D^+_{b'}(p)]^{-1} = -\omega' Q^\textsc{a}_{c'} [\hat{S}_{c'}(\overline{p}_{b'}) - \hat{S}_{c'}(\overline{2p}_{b'})] + \mathcal{O}(\absto{p}{\theta - 1})$. By the same argument adopted above, this coincides with the $\sharp = \textsc{a}$ version of~\eqref{eq:constraints}. 

\paragraph{Proof of~\eqref{eq:contractions}.} If we contract both sides of~\eqref{eq:Pi_explicit} with $ip_\mu$, we obtain
\begin{equation}
\label{eq:Pi_axial_contraction}
i p_\mu \hat{\Pi}^{\mu \nu}_\rff(p) = - \sum_{b'} \frac{\sigma^\textsc{v}_{\nu, b'} D^-_{b'}}{4\pi \abs*{\mathfrak{u}_{b'} \!}\! \mathcal{Z}_{b'}} \left[ \sum_b ip_\mu \sigma^\textsc{a}_{\mu, b} \frac{T_{bb'}(p)}{\mathcal{Z}_b D^+_{b'}(p)} \right] = -\sum_{b'} \frac{\sigma^\textsc{v}_{\nu, b'} D^-_{b'}}{4\pi \abs*{\mathfrak{u}_{b'} \!}\! \mathcal{Z}_{b'}} [-\omega' Q^\textsc{a}_{c'}] + \mathcal{O}(\absto{p}{1 + \theta}),
\end{equation}
where the relation~\eqref{eq:constraints} with $\sharp = \textsc{a}$ has been used in the second equality. The first of~\eqref{eq:contractions} then follows by noting that $D^-_{b'}(p) = -i p_0 - \omega' \abs*{\mathfrak{u}_{b'}} p_1 = -(i, \omega' \abs*{\mathfrak{u}_{b'}}\!)_\alpha p^\alpha$.

The second of~\eqref{eq:contractions} follows from a similar reasoning. This time, however, we preliminarily observe that
\begin{equation}
\label{eq:symmetry_relation}
\begin{aligned}[c]
\frac{D^-_b}{D^+_b \abs*{\mathfrak{u}_b}\!} (T^{-1})_{b'b} = \frac{D^-_b}{D^+_b \abs*{\mathfrak{u}_b}\!} \left( \delta_{b'b} - \frac{\lambda^\rff_{b'b} D^-_{b'}}{4\pi \abs*{\mathfrak{u}_{b'}}\! D^+_{b'}} \right)
& = \frac{D^-_{b'}}{D^+_{b'} \abs*{\mathfrak{u}_{b'}}\!} \delta_{bb'} - \frac{D^-_b}{\abs*{\mathfrak{u}_b}\! D^+_b} \lambda^\rff_{b'b} \frac{D^-_{b'}}{4\pi \abs*{\mathfrak{u}_{b'}}\! D^+_{b'}} \\
& = \left( \delta_{bb'} - \frac{D^-_b}{4\pi \abs*{\mathfrak{u}_b}\! D^+_b} \lambda^\rff_{bb'} \right)  \frac{D^-_{b'}}{\abs*{\mathfrak{u}_{b'}}\! D^+_{b'}} \\
& = (T^{-1})_{bb'} \frac{D^-_{b'}}{\abs*{\mathfrak{u}_{b'}}\! D^+_{b'}}.
\end{aligned}
\end{equation}
If we let $H_{bb'}(p) \equiv \delta_{bb'} D_{b'}^-(p)[D_{b'}^+(p) \abs*{\mathfrak{u}_{b'}}]^{-1}$, the identity~\eqref{eq:symmetry_relation} reads $H (T^\mathrm{t})^{-1} = T^{-1} H$: after multiplying both sides by $T$ and $T^\mathrm{t}$, we obtain $T H = H T^\mathrm{t}$, that is $T_{bb'}(p) D_{b'}^-(p)[D_{b'}^+(p) \abs*{\mathfrak{u}_{b'}}]^{-1} = T_{b'b}(p) D_b^-(p)[D_b^+(p) \abs*{\mathfrak{u}_b}]^{-1}$. Thanks to this relation, we can recast~\eqref{eq:Pi_explicit} into the alternative form
\begin{equation}
\label{eq:Pi_explicit_alternative}
\hat{\Pi}^{\mu \nu}_\rff(p) = - \sum_{bb'} \sigma^\textsc{a}_{\mu, b'} \sigma^\textsc{v}_{\nu, b} \frac{T_{bb'}(p) D^-_{b'}(p)}{4\pi \abs*{\mathfrak{u}_{b'}} \! \mathcal{Z}_{b'} \mathcal{Z}_b D^+_{b'}(p)}.
\end{equation}
Finally, by contracting both sides of~\eqref{eq:Pi_explicit_alternative} with $ip_\nu$ and using~\eqref{eq:constraints}, we obtain
\begin{equation}
\label{eq:Pi_vector_contraction}
i p_\nu \hat{\Pi}^{\mu \nu}_\rff(p) = - \sum_{b'} \frac{\sigma^\textsc{a}_{\mu, b'} D^-_{b'}}{4\pi \abs*{\mathfrak{u}_{b'} \!}\! \mathcal{Z}_{b'}} \left[ \sum_b ip_\nu \sigma^\textsc{v}_{\nu, b} \frac{T_{bb'}(p)}{\mathcal{Z}_b D^+_{b'}(p)} \right] = -\sum_{b'} \frac{\sigma^\textsc{a}_{\mu, b'} D^-_{b'}}{4\pi \abs*{\mathfrak{u}_{b'} \!}\! \mathcal{Z}_{b'}} [- Q^\textsc{v}_{c'}],
\end{equation}
and this proves the second of~\eqref{eq:contractions}.

\paragraph{Proof of~\eqref{eq:M_matrix}.} By plugging the second of~\eqref{eq:contractions} into~\eqref{eq:R_constant_determination} and recalling that $\sigma_{\mu, b}^\textsc{v} = (\mathcal{Z}^\textsc{v}_{0, b}, i\omega \mathcal{Z}^\textsc{v}_{1, b})_\mu$ and $\sigma_{\mu, b}^\textsc{a} = (\omega \mathcal{Z}^\textsc{a}_{0, b}, i \mathcal{Z}^\textsc{a}_{1, b})_\mu$, we readily get
\begin{equation}
\label{eq:vector_contraction_explicit}
i\mathscr{R}^{\mu \nu}(0) = \sum_b (i, \omega \abs*{\mathfrak{u}_b} \!)_\nu \frac{\sigma^\textsc{a}_{\mu, b} Q^\textsc{v}_c}{4\pi \abs*{\mathfrak{u}_b} \! \mathcal{Z}_b} = \sum_b \frac{1}{4\pi \mathcal{Z}_b}
\begin{pmatrix}
\displaystyle
i\omega \frac{\mathcal{Z}^\textsc{a}_{0, b} Q^\textsc{v}_c}{\abs*{\mathfrak{u}_b}\!}
& - \displaystyle \frac{\mathcal{Z}^\textsc{a}_{1, b} Q^\textsc{v}_c}{\abs*{\mathfrak{u}_b}\!} \\[1em]
\mathcal{Z}^\textsc{a}_{0, b} Q^\textsc{v}_c
& i \omega \mathcal{Z}^\textsc{a}_{1, b} Q^\textsc{v}_c
\end{pmatrix}_{\!\! \nu \mu} \equiv \sum_b M_{1, b}^{\nu \mu}.
\end{equation}
Note that the matrices $M_{1, b}$ have been defined so that $\sum_b M_{1, b}$ is equal to the \emph{transpose} of $i \mathscr{R}(0)$. Similarly, the first of~\eqref{eq:contractions} tells us that
\begin{equation}
\label{eq:axial_contraction_explicit}
\begin{aligned}[c]
ip_\alpha \hat{\Pi}^{\alpha \mu}_\rff(p) = -p^\alpha \sum_b (i, \omega \abs*{\mathfrak{u}_b} \!)_\alpha \frac{\sigma^\textsc{v}_{\mu, b} \, \omega Q^\textsc{a}_c}{4\pi \abs*{\mathfrak{u}_b} \! \mathcal{Z}_b} 
& = \sum_b \frac{1}{4\pi \mathcal{Z}_b} \begin{pmatrix}
\displaystyle
-i\omega \frac{\mathcal{Z}^\textsc{v}_{0, b} Q^\textsc{a}_c}{\abs*{\mathfrak{u}_b}\!}
& - \mathcal{Z}^\textsc{v}_{0, b} Q^\textsc{a}_c \\[1em]
\displaystyle
\frac{\mathcal{Z}^\textsc{v}_{1, b} Q^\textsc{a}_c}{\abs*{\mathfrak{u}_b}\!}
& -i \omega \mathcal{Z}^\textsc{v}_{1, b} Q^\textsc{a}_c
\end{pmatrix}_{\!\! \mu \alpha} p_\alpha \\
& \equiv \sum_b M_{2, b}^{\mu \alpha} \, p_\alpha
\end{aligned}
\end{equation}
up to subdominant corrections. Knowing that $ip_\alpha \hat{\Pi}^{\alpha \mu}(p) = ip_\alpha[\hat{\Pi}^{\alpha \mu}_\rff(p) + \mathscr{R}^{\alpha \mu}(0)] + \mathcal{O}(\absto{p}{1 + \theta})$, formulae~\eqref{eq:vector_contraction_explicit} and~\eqref{eq:axial_contraction_explicit} yield
\begin{equation}
ip_\alpha \hat{\Pi}^{\alpha \mu}(p) = \sum_b M_{2, b}^{\mu \alpha} \, p_\alpha + p_\alpha \cdot \sum_b M_{1, b}^{\mu \alpha} + \mathcal{O}(\absto{p}{1 + \theta})= \sum_b (M^{\mu \alpha}_{1, b} + M^{\mu \alpha}_{2, b}) \, p_\alpha + \mathcal{O}(\absto{p}{1 + \theta}).
\end{equation}
The explicit form of the sum $M^{\mu \alpha}_{1, b} + M^{\mu \alpha}_{2, b}$ can be immediately read off from~\eqref{eq:vector_contraction_explicit} and~\eqref{eq:axial_contraction_explicit}, namely
\begin{equation}
M^{\mu \alpha}_{1, b} + M^{\mu \alpha}_{2, b} =
\frac{1}{4\pi\mathcal{Z}_b}
\begin{pmatrix}
\displaystyle i\omega\frac{\mathcal{Z}^\textsc{a}_{0, b} Q^\textsc{v}_c - \mathcal{Z}^\textsc{v}_{0, b} Q^\textsc{a}_c}{\abs*{\mathfrak{u}_b}\!} 
&  & -\mathcal{Z}^\textsc{v}_{0, b} Q^\textsc{a}_c - \displaystyle \frac{\mathcal{Z}^\textsc{a}_{1, b}Q^\textsc{v}_c}{\abs*{\mathfrak{u}_b}\!} \\[10pt]
\displaystyle \frac{\mathcal{Z}^\textsc{v}_{1, b} Q^\textsc{a}_c}{\abs*{\mathfrak{u}_b}\!} + \mathcal{Z}^\textsc{a}_{0, b} Q^\textsc{v}_c
&  & i\omega(\mathcal{Z}^\textsc{a}_{1, b} Q^\textsc{v}_c - \mathcal{Z}^\textsc{v}_{1, b} Q^\textsc{a}_c)
\end{pmatrix}_{\!\! \mu \alpha}
\equiv M_b^{\mu \alpha},
\end{equation}
which coincides with~\eqref{eq:M_matrix}.
\end{document}